\documentclass{aa}  

\usepackage{graphicx}

\usepackage{natbib}
\usepackage{rotating}     
\usepackage{booktabs}     
\usepackage{multirow}     
\usepackage{hhline}       
\usepackage{csvsimple}    
\usepackage{xcolor}
\usepackage[colorlinks,
            linkcolor=blue,
            citecolor=blue,
            urlcolor=blue]{hyperref}
\bibpunct{(}{)}{;}{a}{}{,}

\usepackage{txfonts}

\begin{document}

   \title{Physical characterization
of the local system NGC 1313: A flattening in its chemical abundance by past interactions 
}

   \author{Belén Ojeda-Díaz 
          \inst{1} \fnmsep\thanks{ \email{belen.ojeda@userena.cl} (BOD) } \and
          Sergio Torres-Flores\inst{1} \and
          Daniela E. Olave-Rojas\inst{2} \and
          Ciria Lima-Dias\inst{3} \and
          Paulo J. A. Lago\inst{4} \and
          Claudia Mendes de Oliveira\inst{5} \and
          Philippe Amram\inst{6} \and
          Bruno C. Quint\inst{7}
          }

   \institute{Departamento de Astronomía, Universidad de La Serena, Avda. Raúl Bitrán 1305, La Serena, Chile.
         \and
             Departamento de Tecnologías Industriales, Facultad de Ingeniería, Universidad de Talca, Los Niches km 1, Curicó, Chile.
        \and
            Instituto Multidisciplinario de Investigación y Postgrado, Universidad de La Serena, Avda. Raúl Bitrán 1305, La Serena, Chile.
        \and
            Vera C. Rubin Observatory, Avda. Juan Cisternas 1500, La Serena, Chile.
        \and 
            Departamento de Astronomia, Instituto de Astronomia, Geof\'isica e Ci\^{e}ncias Atmosf\'ericas da Universidade de Sa\~{o} Paulo, Cidade Universitaria, 05508-900 Sa\~{o} Paulo, SP, Brazil.
        \and 
            Aix-Marseille Univ., CNRS, CNES, LAM, 38 rue Frédéric Joliot Curie, 13338 France
        \and
            Vera C. Rubin Observatory Project Office, 950 N. Cherry Ave., Tucson, AZ 85719, USA
             }

   \date{Received September 15, 1996; accepted March 16, 1997}


  \abstract
   {Interacting galaxies provide unique information on morphological transformation, enhanced star formation, and chemical evolution, and thus contribute to understanding the complex evolution of galaxies.}
   {We investigated the local interacting system NGC 1313 by analyzing its main physical and kinematical properties to understand how the interaction has influenced the evolutionary state of the galaxy.}
   {We used multi‑slit GMOS‑S spectroscopy to study 19 regions across the galaxy, encompassing its main body and the complex southwest region. We derived oxygen abundances using the N2 method and computed their chemical gradient. We derived electron densities using the S[\textsc{ii}] ratio, ages from the EW(H$\alpha$), and stellar masses of the regions from DR10 Legacy Survey images. We used H$\alpha$ Fabry-Perot data to analyze the kinematics of the systems and search for signs of past interactions.}
   {The Baldwin–Phillips–Terlevich (BPT) and EW$_{H\alpha}$ versus N[\textsc{ii}]/H$\alpha$ (WHAN) 
diagnostics confirm photoionization by star formation. The galaxy has a low oxygen abundance (8.0$<$12+log(O/H)$<$8.2), with a mainly flat oxygen abundance gradient, suggesting gas mixing processes. Electron densities span $n_e$$\sim<10$ to 142 cm$^{-3}$. The blue and red Wolf-Rayet bumps detected in two regions corroborate a young population. We find ages ranging from $2.7$ Myr to $6.0$ Myr. The velocity field shows complex kinematics in the northern region of NGC 1313, characterized by asymmetric line profiles.}
   {NGC 1313 thus provides an ideal laboratory for studying how minor interactions affect star formation, chemical enrichment, and the kinematics of low-mass barred spiral galaxies}


\authorrunning{Ojeda-Díaz et al.}
\titlerunning{Star formation in NGC 1313}

   \maketitle

%

\section{Introduction}

Gravitational interactions significantly impact galaxy morphology (\citealt{toomre&toomre1972}, \citealt{barnes&hernquist1992}), star formation rates (\citealt{kennicutt1998}, \citealt{ellison+2008}, \citealt{patton+2013,patton+2020}), and chemical content (\citealt{kewley+2010}, \citealt{bresolin+2012}, \citealt{torres-flores+2014}, \citealt{pan+2025}). Studying the nature and occurrence of interacting systems is a key element in understanding galaxy formation and evolution. This is supported by the hierarchical scenario of galaxy formation and evolution, in which less massive galaxies merge to form increasingly more massive galaxies \citep{toomre&toomre1972,papovich+2015}. Consequently, studying the interaction between galaxies in local systems provides important information about the processes that were prevalent in a distant Universe, particularly regarding morphology. We classify galaxy pair mergers by mass ratio: $>$1:4 for major mergers and $\lesssim$1:4 for minor mergers (\citealt{kaviraj+2014}). Each regime yields distinct evolutionary consequences. Major mergers are violent and destructive, as both progenitors are destroyed and merge to form a galaxy that resembles neither.  Conversely, a minor merger results in a galaxy similar to the more massive progenitor.

The chemical distribution is a useful tool for determining whether a galaxy has undergone an interaction, since observations (\citealt{kewley+2010}) and simulations (\citealt{perez+2011}; \citealt{tapia-contreras+2025}) have shown that a flat chemical gradient represents a mixing of gasses due to gravitational interactions. This can be caused by inflows of pristine gas that dilute the central abundance \citep{rupke+2010a} or by outflows of enriched gas towards the galaxy's outskirts. However, recent studies show that this gradient can be flat, inverted, or even steeper than before the interaction, depending on the type of merger \citep{pan+2025}. Analyzing the radial chemical distribution in these systems provides unique information about the influence of the interaction on the physical properties of the galaxies involved.

The local system NGC 1313 exhibits a perturbed morphology with asymmetric arms composed of star-forming regions and substructures to the southwest. This motivated us to explore the origin of these strong signs of interaction. This system lacks an obvious companion and does not belong to a galaxy group, which makes it an interesting object of study in the context of galaxy formation and evolution.

NGC 1313 is a nearby  barred galaxy of type SB(s)d (z=0.001568, \citealt{ohlson+2024}) located at a distance of 4.19 Mpc, as estimated via the top of the red giant branch (TRGB; \citealt{sabbi+2018}). The central position of NGC 1313 is right ascension (R.A.)  03:18:16.050s and declination (Dec) -66:29:53.70s in J2000 equatorial coordinates. Due to its moderate stellar mass ($\sim$2.6$\times$10$^9$ M$_\odot$; \citealt{calzetti+2015}) and irregular appearance, it is often classified as a Magellanic barred spiral. NGC 1313 appears to be in a transitional stage between dwarf irregulars and typical spiral disks, reminiscent of the Large Magellanic Cloud (LMC)  \citet{deVaucouleurs1963}. Despite the lack of a major companion, several clues hint that NGC 1313 may have experienced a minor interaction in the past.  \citet{peters+1994} first proposed this scenario when analyzing the distribution of neutral hydrogen and its velocity field. They found that, in addition to the typical emission of a galaxy of this type, an ongoing tidal interaction between NGC 1313 and a satellite galaxy in the southwest of the system resulted in a hydrogen loop. This idea was later revisited by \citet{silva-villa+2012}, who used resolved stellar populations and found a burst of star formation in the southwestern region of the galaxy in its star formation history (SFH). However, these authors found no evidence of interactions in other regions of the galaxy, suggesting that the interacting object was a moderate-sized galaxy that did not trigger a global burst of star formation.

Previous studies have also analyzed the chemical distribution in NGC 1313. Using multifiber data, \cite{walsh+1997} found that this system displays a relatively flat metal distribution. These authors indicated that NGC 1313 is the most massive barred galaxy presenting no gradient distribution. In contrast, \cite{Hernandez2022} used spectroscopic data of five massive young star clusters in this system and found a steeper , negative metallicity gradient for this galaxy and argue that the discrepancy with \cite{walsh+1997} arises from the different metallicity calibrators used. The metal distribution in this galaxy offers an opportunity to explore scenarios that can explain flat or even inverted metallicity gradients (e.g., \citealt{kewley+2010}), where galaxy-galaxy interactions appear to play a relevant role in the mixture of enriched and less enriched gasses (e.g., \citealt{torres-flores+2014}, \citealt{olave-rojas+2015}). For this reason, this work explores the interacting system NGC 1313, where previous studies found a homogeneous distribution of gas-phase oxygen abundance \citep{walsh+1997}. We focus on the physical properties of 19 sources located across the disk of NGC 1313 and in regions located in the southwest of the galaxy. The southwest region is of particular interest because previous studies found enhanced star formation activity \citep{silva-villa+2012}, which can be triggered by tidal interactions (\citealt{peters+1994}). This work contributes to our understanding of star formation and chemical distribution in interacting systems.

This paper is organized as follows. In Section \ref{sec:data} we present the spectroscopic data and its reduction. In Section \ref{sec:analysis} we describe the data analysis, including emission-line measurements, extinction corrections, and estimations of physical properties such as electron density, oxygen abundances, abundance gradients, stellar masses, and ages. We present the results in Section \ref{sec:results}. We discuss our results in Section \ref{sec:discussion} and provide general conclusions in Section \ref{sec:summary}. Throughout this paper, we propagate uncertainties using the \textsc{uncertainties} \textsc{Python} library. This library uses linear error propagation theory, automatically calculating derivatives and analytically propagating them to the results. Consequently, the uncertainty associated with each result depends on the applied mathematical operation and incorporates the propagated uncertainties from all measurements and parameters involved in the governing equation.

\begin{figure}
    \centering
    \includegraphics[width=1.05\linewidth]{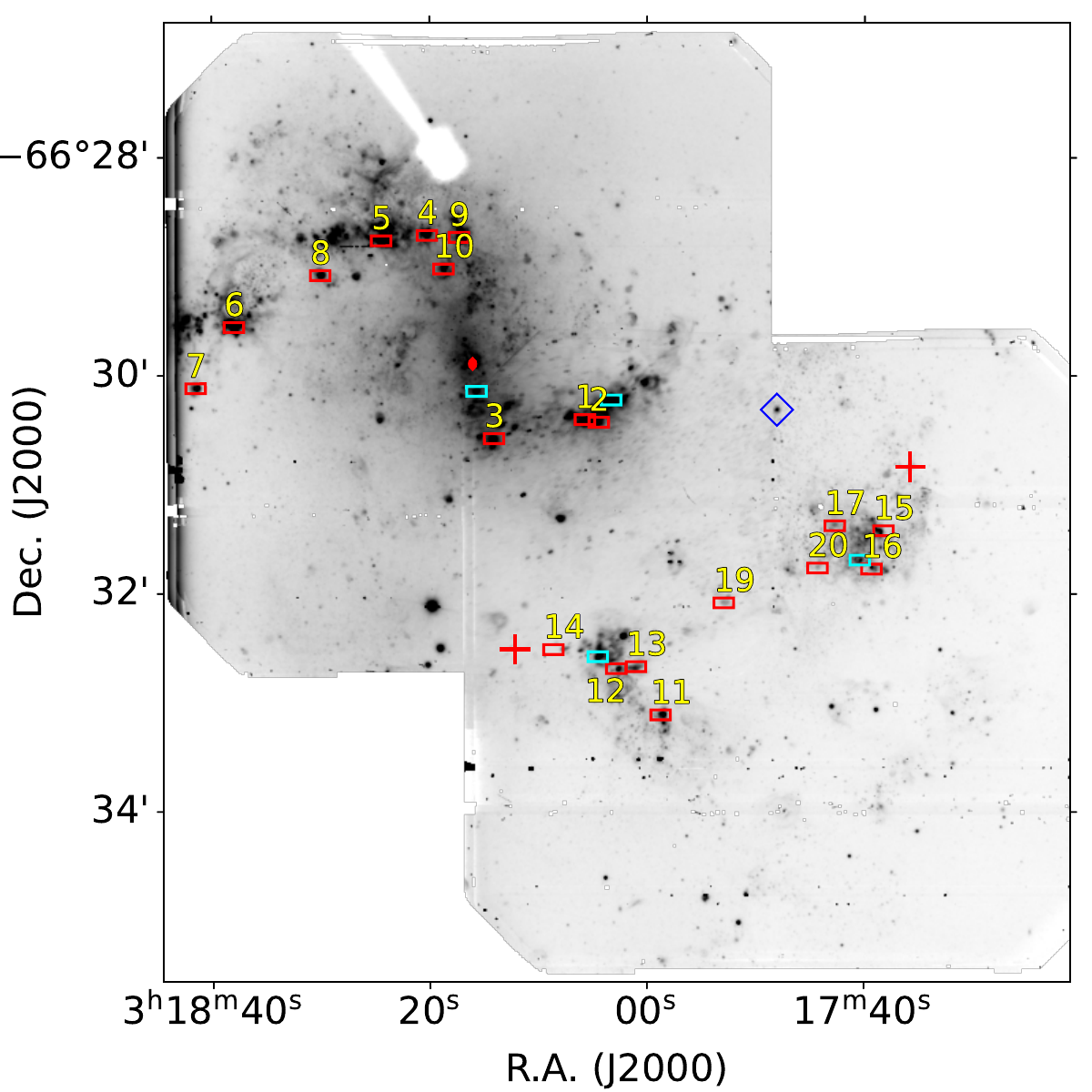}
    \includegraphics[width=1.05\linewidth]{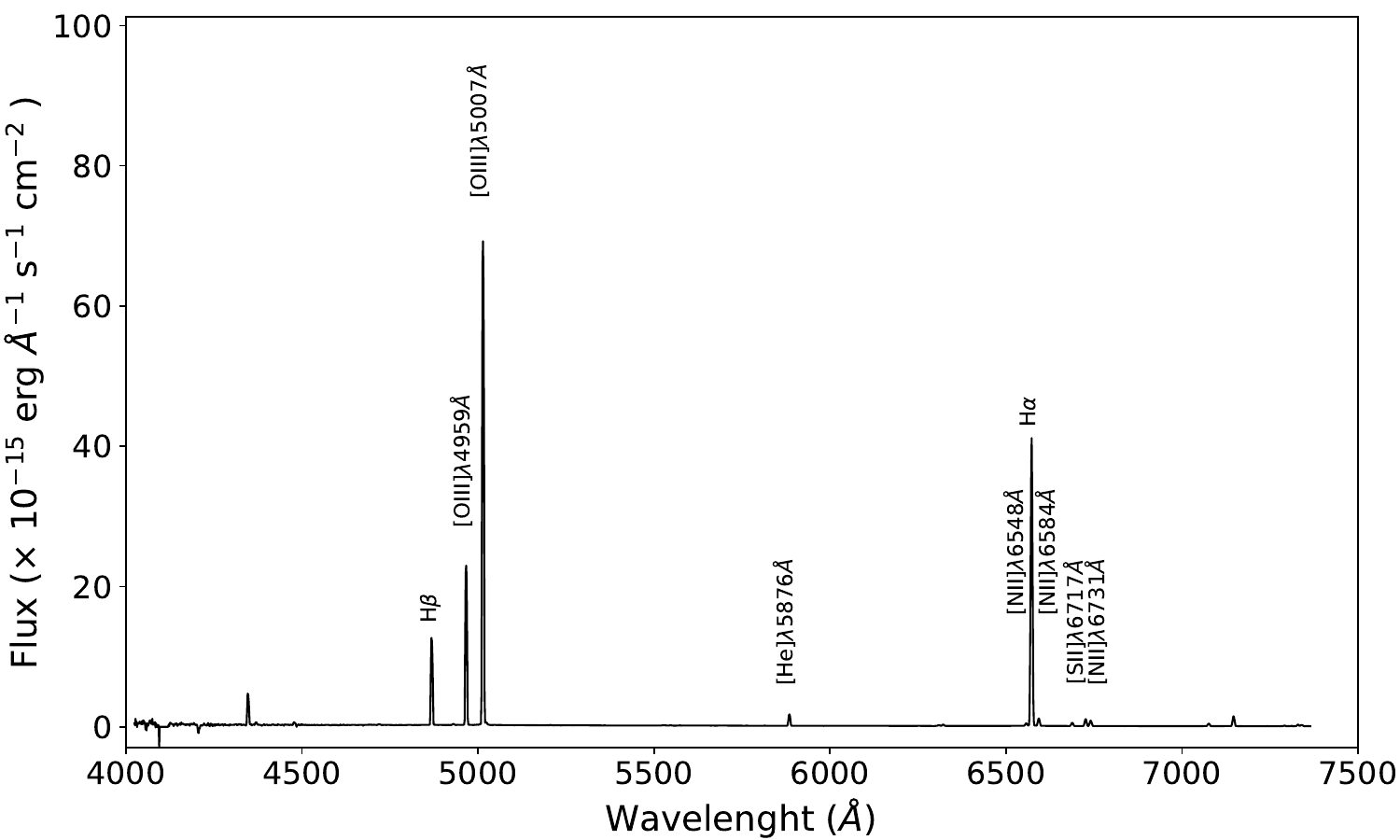}
    \caption{Top panel: H$\alpha$ images of the two Gemini Multi-Object Spectrograph (GMOS) fields observed in NGC 1313. Field 1 (upper left) covers the main body of the galaxy, while field 2 (lower right) covers the star-forming clouds in the southwest. Red rectangles indicate the slit positions for the observed H$\textsc{ii}$ regions, with their IDs labeled in yellow. Red crosses indicate regions without detected emission. Cyan rectangles indicate regions where emission was detected but not analyzed (see section \ref{sec:data}). The blue diamond indicates a star cluster observed by its H$\alpha$ emission. The red circle indicates the center of NGC 1313 from the NASA/IPAC Extragalactic Database (NED). North is up and east is to the left. Bottom panel: Spectrum of region \#1 observed in NGC 1313, corrected for extinction using the law of \citet{calzetti+2000}. Emission lines characteristics of a star-forming region are labeled.}
    \label{fig:ngc1313_slits_center}
\end{figure}

\section{Data} \label{sec:data}

We observed NGC 1313 with the Gemini Multi-Object Spectrograph (GMOS) at the Gemini South telescope during the second semester of 2019, under program GS-2019B-Q-103 (PI: Paulo J. A. Lago). To characterize the chemical composition of the ionized gas in an extended area, we observed two GMOS fields of 5.5 $\times$ 5.5 arcmin$^2$ each. One field was centered on the main body of the galaxy (field 1) and the other  includes the southwest region of the system (field 2). To create the GMOS spectroscopic mask, we used H$\alpha$ observations of NGC 1313 to identify candidate star-forming regions. 

During the observations, the seeing was typically 0.7 arcsec, so we adopted a slit width of 0.75 arcsec to ensure proper sampling while maximizing throughput and maintaining spectral resolution. The instrument position angle was 90.1º. We made two observations for each field using the R400\_G5325 grating, with exposure times of 1050 s and 1079 s for fields 1 and 2, respectively, and centered at wavelengths of 5200 \AA\ and 5300 \AA\ for each field. In all cases, the wavelength coverage ranged from 4130 - 7645 \AA, encompassing the emission lines needed to estimate the physical properties of star-forming regions. The spectral resolution at H$\alpha$ was R$\sim$1300, with $\Delta\lambda$ $\sim5$\AA. 

We reduced the GMOS spectroscopic data using the \textsc{gemini} package (v1.14) in \textsc{iraf} (v2.16). All science and calibration frames were corrected by overscan, bias and flat-field using the tasks \textsc{gbias, gsflat,} and \textsc{gsreduce}. We performed wavelength calibration with the \textsc{gswavelength} and \textsc{gstransform} routines, obtaining an rms of 0.28 from the CuAr lamp calibration. We removed cosmic rays using the \textsc{lacos\_spec} task of \citet{vanDokkum2001} through the routine \textsc{gemcrspec}. We extracted the spectra from the mask using the \textsc{gscut} task and combined them using the \textsc{gemcombine} task. We then used the \textsc{gsextract} routine to extract spectra from the sky. We reduced the spectrum of the observed standard star LTT3218 in the same way as the galaxy data in order to derive the sensitivity function with \textsc{gsstandard}. Finally, we flux-calibrated all spectra using the \textsc{gscalibrate} task.

Figure \ref{fig:ngc1313_slits_center} shows an H$\alpha$ image of the two fields observed in this work. The figure includes the observed and analyzed regions (with their identifications), those where no emission was detected, and the regions that showed emission but where the H$\alpha$ emission was so intense that it contaminated the entire slit, preventing spectral extraction.
One region selected for its H$\alpha$ emission turned out to be a young star cluster and was therefore not analyzed. \citet{Hernandez2022} named this cluster ``NGC 1313-379.'' The bottom panel of Figure \ref{fig:ngc1313_slits_center} shows an example of the typical observed spectrum in NGC 1313.

\subsection{Complementary optical images}

To complement the spectroscopic data, we used 
$i$-band images from the DR10 Legacy Survey\footnote{\url{https://www.legacysurvey.org/dr10/}} \citep{dey+2019}, observed with the Dark Energy Camera (DECam; \citealt{flaugher+2015}). We selected an area of 8 arcmin centered on the coordinates of NGC 1313, from the NASA/IPAC Extragalactic Database (NED). This region covers the same area as the two fields observed with GMOS-S (main body and southwestern structures). The DR10 images have a pixel scale of 0.262" per pixel. The zero-point value in the Legacy Survey images is fixed for all bands ($zp = 22.5$). This is because in DR10 object brightness is stored as linear fluxes in units of nanomaggies. Furthermore, the sky is already subtracted in the DR10 images, so the integrated flux in the region is considered intrinsic to the source. Finally, we used weight images for the same field in the $g$ and $i$ filters to estimate the flux uncertainties for each region. These images contain the error in the flux measurement for each pixel in the observed field.

\subsection{Complementary SAM-Fabry-Perot observations}
\label{fabry_perot}

We observed NGC 1313 with the Southern Astrophysical Research (SOAR) Adaptive Module-Fabry-Perot (SAM-FP) instrument \citep{2017MendesdeOliveira} at the SOAR telescope. This configuration uses SAM, the ground layer adaptive optics instrument \citep{2008Tokovinin}, which improves image quality. For the Fabry-Perot observations, we used an interference order of p=609 at H$\alpha$. In these observations, the free spectral range (FSR) has an amplitude of  490 km s$^{-1}$, which we scan with 38 channels. The resolution power of the observation was R$\simeq$11500. We performed wavelength calibrations by scanning the Ne 6598.95 {\AA} line. \cite{2017MendesdeOliveira} provide a detailed description of the SAM-FP instrument. The SAM-FP covers a field of view of 3'$\times$3'. For NGC 1313, we observed two pointings to cover the main body of the system, which corresponds mainly to field 1 shown in Figure \ref{fig:ngc1313_slits_center}.

We corrected SAM-FP data for flat field, dark images, and bias using the standard routines provided in {\sc iraf}. We performed phase map correction and OH night-sky line subtraction with the {\sc ADHOCw} software \footnote{The ADHOCw software package (Boulesteix 1993, Reference Manual, Observatoire de Marseille)}, following the procedure described by \cite{amram+1996}. Finally, we produced a 3D H$\alpha$ data cube, which allowed us to map the kinematics of the warm ionized gas.

\begin{table*}[h!]
    \centering
    \scriptsize
    \caption{Coordinates of the star-forming regions observed in NGC 1313, together with the emission-line fluxes and associated stellar color excess values.}
    \label{tab:flujos}
    \begin{tabular}{cccccccccc}
         \hline
         & & & & & & & & \\ 
          \multirow{2}{*}{ID} & \multicolumn{1}{c}{R.A.} & \multicolumn{1}{c}{Dec} & \multirow{2}{*}{E(B-V)$_s$} & \multicolumn{1}{c}{H$\beta$ $\lambda$4861} & \multicolumn{1}{c}{[O$\textsc{iii}$] $\lambda$5007} & \multicolumn{1}{c}{H$\alpha$ $\lambda$6563} & \multicolumn{1}{c}{[N$\textsc{ii}$] $\lambda$6584} & \multicolumn{1}{c}{[S$\textsc{ii}$] $\lambda$6716} & \multicolumn{1}{c}{[S$\textsc{ii}$] $\lambda$6731} \\
         & J2000 & J2000 & & & & & & \\
         & & & & & & & & &\\ 
         \hhline{----------}
         & & & & & & & & &\\ 
         \begin{filecontents*}{tablas/measurement_emission_lines_ebv_oficial_newnumber.csv}
ID,ID_MASK,RA,DEC,H_BETA_FLUX,H_BETA_FLUX_ERR,OIII_5007_FLUX,OIII_5007_FLUX_ERR,H_ALPHA_FLUX,H_ALPHA_FLUX_ERR,NII_6584_FLUX,NII_6584_FLUX_ERR,SII_6716_FLUX,SII_6716_FLUX_ERR,SII_6731_FLUX,SII_6731_FLUX_ERR,EBV,EBV_ERR
1,1_mask1,03:18:05.640,-66:30:24.69,76.522,2.164,427.052,16.032,257.488,1.858,7.161,1.894,6.448,0.097,4.958,0.087,0.109,0.013
2,3_mask1,03:18:04.375,-66:30:26.06,14.925,0.593,74.586,3.721,46.165,0.470,1.633,0.447,1.927,0.025,1.433,0.022,0.054,0.016
3,11_mask1,03:18:14.078,-66:30:34.99,11.544,0.493,34.509,1.399,38.739,0.349,2.523,0.328,2.489,0.096,1.584,0.074,0.107,0.017
4,13_mask1,03:18:20.213,-66:28:42.79,27.353,0.721,119.359,4.474,83.540,1.724,3.299,1.744,3.084,0.039,2.358,0.034,0.044,0.013
5,15_mask1,03:18:24.470,-66:28:45.81,41.371,1.183,110.078,2.769,142.615,3.538,7.629,3.140,6.803,0.078,5.175,0.071,0.126,0.016
6,17_mask1,03:18:38.021,-66:29:32.76,26.796,0.970,124.782,3.332,80.038,1.262,3.351,1.233,4.614,0.063,3.507,0.053,0.028,0.015
7,21_mask1,03:18:41.431,-66:30:07.14,9.753,0.311,26.238,0.894,30.152,0.318,1.886,0.307,1.719,0.035,1.212,0.029,0.054,0.013
8,23_mask1,03:18:30.046,-66:29:05.04,6.794,0.350,29.676,0.994,21.257,0.134,0.721,0.123,0.718,0.010,0.506,0.008,0.059,0.019
9,25_mask1,03:18:17.155,-66:28:44.19,8.345,0.284,14.359,0.654,31.093,0.718,0.792,0.213,2.075,0.056,1.627,0.049,0.177,0.020
10,27_mask1,03:18:18.684,-66:29:01.14,3.428,0.130,10.078,0.383,10.281,0.104,0.543,0.095,0.461,0.020,0.273,0.016,0.033,0.014
11,4_mask2,03:17:58.563,-66:33:06.30,10.556,0.435,26.640,1.313,32.019,0.389,2.029,0.384,2.276,0.037,1.616,0.031,0.040,0.016
12,8_mask2,03:18:02.691,-66:32:41.67,5.613,0.221,17.461,1.013,17.434,0.203,0.735,0.197,0.751,0.016,0.530,0.013,0.057,0.015
13,10_mask2,03:18:01.248,-66:32:41.09,1.570,0.102,2.894,0.132,4.562,0.049,0.286,0.046,0.395,0.010,0.253,0.008,0.009,0.023
14,12_mask2,03:18:07.790,-66:32:30.76,0.962,0.054,3.684,0.194,2.809,0.042,0.165,0.042,0.189,0.006,0.127,0.004,0.012,0.022
15,15_mask2,03:17:38.419,-66:31:25.36,5.913,0.164,17.750,0.766,19.285,0.358,0.909,0.394,1.115,0.018,0.765,0.014,0.088,0.014
16,17_mask2,03:17:39.384,-66:31:45.77,3.832,0.228,12.797,0.627,12.506,0.193,0.476,0.191,0.577,0.019,0.426,0.016,0.087,0.023
17,23_mask2,03:17:42.749,-66:31:21.99,0.948,0.067,2.168,0.110,2.780,0.057,0.189,0.052,0.293,0.009,0.214,0.008,0.016,0.028
19,27_mask2,03:17:52.915,-66:32:05.00,0.558,0.083,3.781,0.137,1.629,0.027,0.118,0.024,0.212,0.013,0.147,0.011,0.008,0.060
20,29_mask2,03:17:44.344,-66:31:45.36,0.452,0.117,0.445,0.064,1.400,0.020,0.082,0.016,0.181,0.012,0.138,0.011,0.040,0.101
\end{filecontents*}
\csvreader[late after line=\\, late after last line=\\\hline]{tablas/measurement_emission_lines_ebv_oficial_newnumber.csv}{}{
            \csvcoli & \csvcoliii & \csvcoliv & \csvcolxvii $\ \pm$ \csvcolxviii & \csvcolv $\ \pm$ \csvcolvi & \csvcolvii $\ \pm$ \csvcolviii & \csvcolix $\ \pm$ \csvcolx & \csvcolxi $\ \pm$ \csvcolxii & \csvcolxiii $\ \pm$ \csvcolxiv & \csvcolxv $\ \pm$ \csvcolxvi}
    \end{tabular}\vspace{2mm}
    \tablefoot{All fluxes are in units of [$\times$ 10$^{-15}$ erg s$^{-1}$ cm$^{-2}$ \AA$^{-1}$] and corrected by extinction.}
    \begin{minipage}{\textwidth}
    \raggedright
    \end{minipage}
\end{table*}

\section{Analysis}\label{sec:analysis}

\subsection{Extinction and emission-line measurements}
Star-forming regions are obscured by dust. In our analysis, we used the extinction law proposed by \citet{calzetti+2000} to correct for extinction, as it is designed for starburst galaxies. 
We implemented this correction through the \textsc{extinction} library in \textsc{Python}. For this purpose, we first estimated the gaseous color excess from the Balmer decrement, assuming  H$\alpha$/H$\beta$=2.86 for an electron temperature of $T_e=10 000\ K$ and an electron density of $n_e=100\ cm^{-3}$ (under Case B recombination; \citealt{OF2006}), following the prescriptions of \citet{dominguez+2013}. We used this gaseous color excess to obtain the stellar color excess in the regions through the expression of \citet{calzetti+2000}: E(B-V)$_{stellar}$ = 0.44 E(B-V)$_{gas}$.  We then applied these derived stellar color excesses, which range from 0.008 to 0.17, to correct each region for extinction. Table \ref{tab:flujos} lists the ID, R.A.,  and Dec of the sources. In the fourth column we list the stellar color excess for each observed region. We propagated the uncertainties in the stellar color excess by accounting for the uncertainties in the H$\alpha$ and H$\beta$ emission-line measurements and in the coefficients used to convert gaseous to stellar color excess.

For each spectrum, we estimated the emission-line fluxes of H$\beta$, [O\textsc{iii}] $\lambda$5007 \AA, H$\alpha$, [N\textsc{ii}] $\lambda$6584 \AA, and [S\textsc{ii}] $\lambda\lambda$6716/6731 \AA. Given the high signal-to-noise ratio of the spectra (mean S/N $\sim$200), we employed an automated Gaussian fitting procedure using the \textsc{lime} library developed by \citet{fernandez+2024}. This code fits Gaussian profiles to emission lines by defining spectral bands, which include the emission line and two adjacent continuum regions free of features. The continuum associated with the stellar component is subtracted to obtain the intrinsic line flux. The library returns the Gaussian flux and its associated uncertainty, which corresponds to the 1$\sigma$ standard error derived via a least-squares minimization algorithm using the \textsc{lmfit} package. In Table \ref{tab:flujos} we list the extinction corrected fluxes of the measured emission lines in NGC 1313 (columns five to ten).

\subsection{Oxygen abundances}

A fundamental physical property in the study of galaxy evolution, particularly in interacting systems, is their chemical content. It is well known that galaxy interactions can alter the distribution of metals. Therefore, analyzing the behavior of chemical abundances becomes a fundamental step in the study of NGC 1313. We focused on the analysis of the oxygen abundance. Since we did not detect [O\textsc{iii}] $\lambda$4363 \AA\ in any spectra, we did not apply the direct method to estimate abundances. Instead, we adopted semi-empirical calibrations, specifically the N2 method following the calibrations proposed by \citet{marino+2013}, which exhibits intrinsic dispersions of 0.16 dex. We note that we did not use the O3N2 method given its valid range. We obtained uncertainties in oxygen abundances by considering errors in the estimated emission-line fluxes and in the constant calibrator parameters, and then adding the intrinsic quadrature calibrator dispersion.

To analyze the radial distribution of oxygen abundances, we estimated the deprojected distances to each observed region using the method described by \citet{scarano+2008}. The inclination and position angle of NGC 1313 are $i=(48\pm3)$° and $PA=(0\pm3)$°, respectively, both taken from \citet{ryder+1995}. Thus, we estimated uncertainties of deprojected distances by propagating the $\pm 3$° uncertainties.

In Table \ref{tab:prop_fisicas} we list the identification of each source, the estimated deprojected distances from the center of the galaxy to each observed region, the N2 indexes and the oxygen abundances derived from the N2 method.

\subsection{Electron densities}

We estimated the electron density ($n_e$) in each region using the [S\textsc{ii}] $\lambda\lambda$6716/6731 \AA\ line ratio (RS2) through the \textsc{temden} task of the \textsc{nebular} package of \textsc{stsdas} within \textsc{iraf}. This task calculates the electron density using the five-level program for ions, \textsc{FIVEL}, developed by \citet{derobertis+1987}. To perform this calculation, an electron temperature of T = 10000 K is assumed, which is a typical value for star-forming regions \citep{OF2006}. We show the values obtained for the RS2 ratios and their uncertainties, along with the electron densities of the studied regions, in the last columns of Table \ref{tab:prop_fisicas}. We propagated the uncertainty in the RS2 ratio using the errors in the estimation of the [S\textsc{ii}] doublet flux.

\subsection{Ages and H$\alpha$ equivalent width} 
\label{sec:ages}

We used the equivalent width of the H$\alpha$ emission line, EW(H$\alpha$),  to date our emission-line sources. In this case, H$\alpha$ emission traces ages on the order of a few million years \citep{kroupa+2024}. 
To estimate the ages of the H$\textsc{ii}$ regions, we interpolated the observed EW(H$\alpha$) with the \textsc{starburst99} models (SB99; \citealt{leitherer+1999}). These models consider star formation histories: an instantaneous burst of star formation (also called a single stellar population) or continuous star formation at a constant rate. Additionally, they considered three different initial mass functions (IMFs) in a mass range of 1 to 100 M$_\odot$, along with five different metallicity values (Z=0.04, 0.02, 0.008, 0.004, and 0.001). The models cover ages from 10$^6$ to 10$^9$ years. \citet{leitherer+1999} set the total mass of the simple (or instantaneous) stellar population at 10$^6$ M$_\odot$. These values were chosen to produce properties typical of the observed star-forming regions in galaxies. 

We interpolated the EW(H$\alpha$) values from the model with the observed EW(H$\alpha$) values using the \texttt{Interpolate} subpackage of the \texttt{SciPy} library in \texttt{Python} 3.12.
We obtained the EW(H$\alpha$) used in this analysis with the \textsc{lime} library, which calculates uncertainties using Monte Carlo propagation. Subsequently, we derived the age uncertainty from the confidence interval based on the uncertainties associated with the measurement of EW(H$\alpha$). In this way, we derived an age range for each region, including upper and lower limits. We list the ages and their uncertainties for all star-forming regions in NGC 1313  in the last column of Table \ref{tab:prop_fisicas}. 

\subsection{Stellar masses through photometric data}

In addition to nebular properties, we estimated the stellar masses of each region. To this end, we assumed that the total mass in each H{\sc ii} region is largely traced by its luminosity. In practice, to estimate the stellar masses, we assumed a mass-to-light (M/L) ratio derived from the absolute $i$-band magnitude equal to 0.8 (\citealt{2000McGaugh}, \citealt{2004Portinari}). It is important to note that we cannot use color–mass calibrations (e.g., \citealt{taylor+2011}) to estimate stellar masses of individual H{\sc ii} regions, as they were developed for integrated galaxy light with old stellar populations, which may not be the case for very young stellar populations in H{\sc ii} regions.

Since the star-forming regions are not point-like sources, the circular area around the aperture is contaminated by emission from the underlying stellar population of NGC 1313. Therefore, we performed fixed-aperture photometry  on the Legacy images using the \textsc{photutils} library in \textsc{python}. We adopted a fixed aperture of 2.62" (10 pixels, given a pixel scale of 0.262" per pixel). We selected this value by visual inspection considering the spatial size of the slit and the region from which the spectrum was extracted. We corrected the obtained magnitudes for extinction using the extinction law of \citet{calzetti+2000}.

Table \ref{tab:mass} lists the apparent magnitudes in the $g$ and $i$ bands, the absolute magnitude in the $i$ band, and the masses derived for the 19 star-forming regions. We propagated uncertainties in the stellar masses by accounting for the uncertainties in the magnitude estimates.

\subsection{Kinematic maps as evidence of interaction events}

We used the SAM-FP H$\alpha$ data cube to analyze the kinematics of the warm ionized gas in the system. We used the \textsc{ADHOCw} software to derive the H$\alpha$ monochromatic map of NGC 1313 and its velocity field. These maps provide valuable information on the kinematic behavior of interacting galaxies, particularly when the spectral resolution allows analysis of the H$\alpha$ emission-line profiles, where multiple components may indicate the existence of gas flows and noncircular motions. Indeed, previous analyses of interacting galaxies have revealed complex kinematics, as in the case of the compact galaxy group HCG 31 (\citealt{Amram2007}, \citealt{GomezEspinoza2023}). \cite{Amram2007} found multiple emission-line profiles in the central region of HCG 31, suggesting that this system is in a pre-merger state.  \cite{munoz-elgueta+2018} present another example of using Fabry-Perot data to study interacting galaxies. These authors analyzed NGC 4656 and its candidate tidal dwarf galaxy. The high resolution of the data reveals noncircular and counter-rotating motions at small-scale, suggesting an interaction between NGC 4656 and its satellite. Therefore, kinematic data can help us disentangle the existence of previous interaction episodes in NGC 1313.

\section{Results}
\label{sec:results}

\subsection{Ionization mechanism: Optical diagnostics diagrams}

\begin{figure}[h!]
    \centering
    \includegraphics[width=1\linewidth]{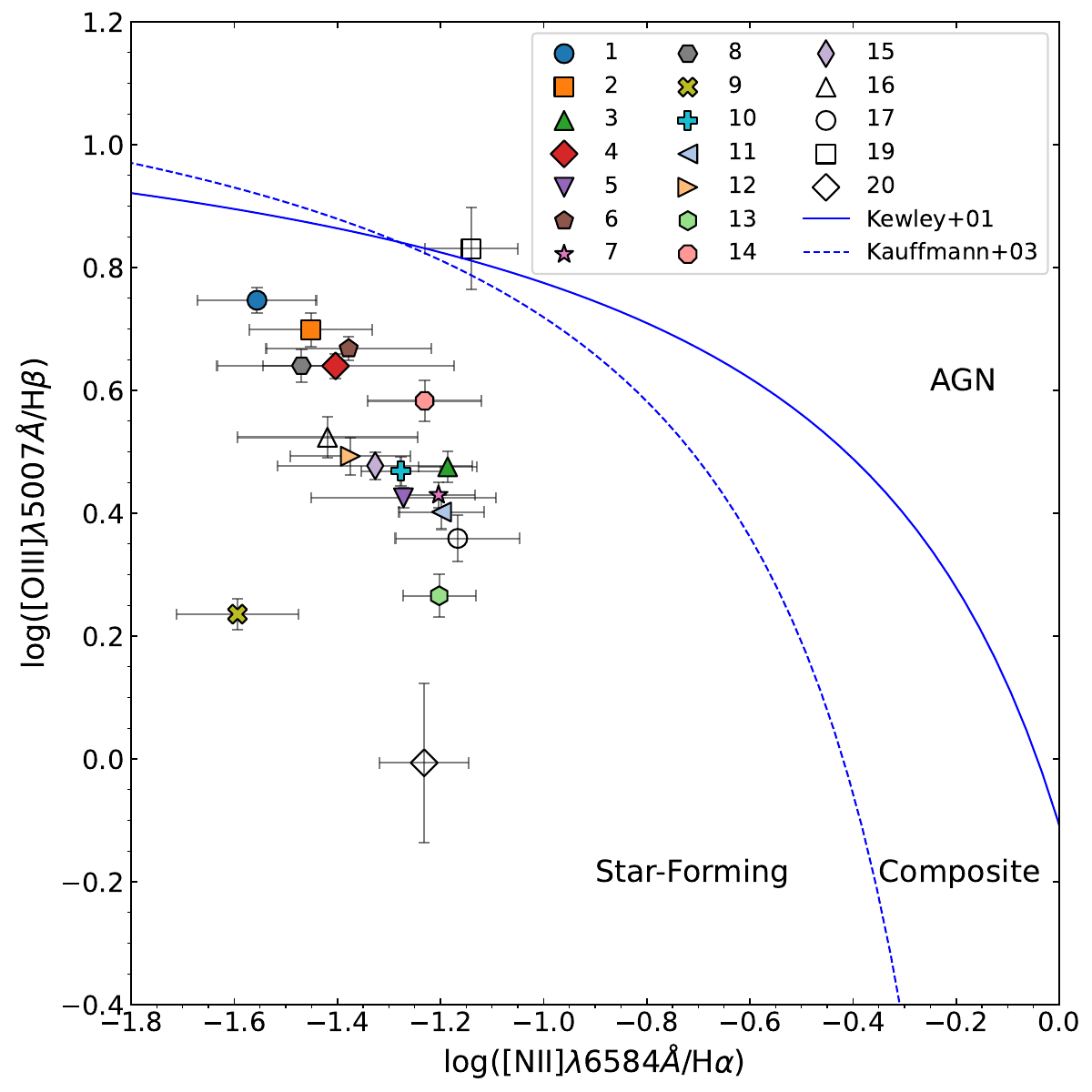}\\
    \includegraphics[width=1\linewidth]{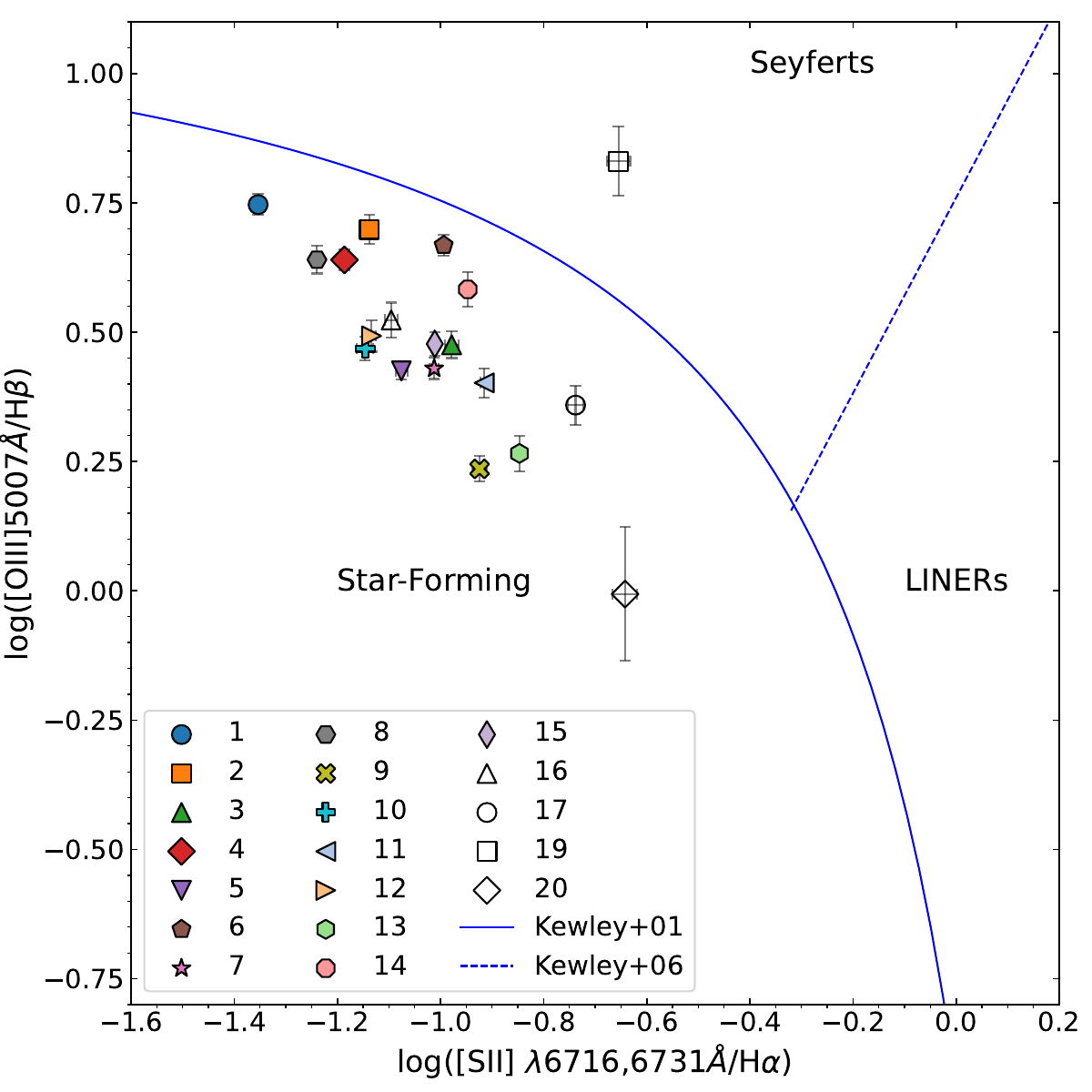}
    \caption{Baldwin–Phillips–Terlevich (BPT) diagram \citep{bpt1981} for the 19 star-forming regions of NGC 1313. Top panel: [O\textsc{iii}]/H$\beta$ versus [N\textsc{ii}]/H$\alpha$. Bottom panel: [O\textsc{iii}]/H$\beta$ versus [S\textsc{ii}]/H$\alpha$. In both panels, the solid blue line represents the limit proposed by \citet{kewley+2001}. In the top panel, the dashed blue line represents the limit of \citet{kauffmann+2003} to differentiate between AGNs and star formation. The dashed blue line in the bottom panel represents the limit of \citet{kewley+2006} to separate Seyferts from low-ionization nuclear emission-line regions (LINERs).}
    \label{fig:bpt_NII}
\end{figure}

\begin{figure}
\centering
\includegraphics[width=1\linewidth]{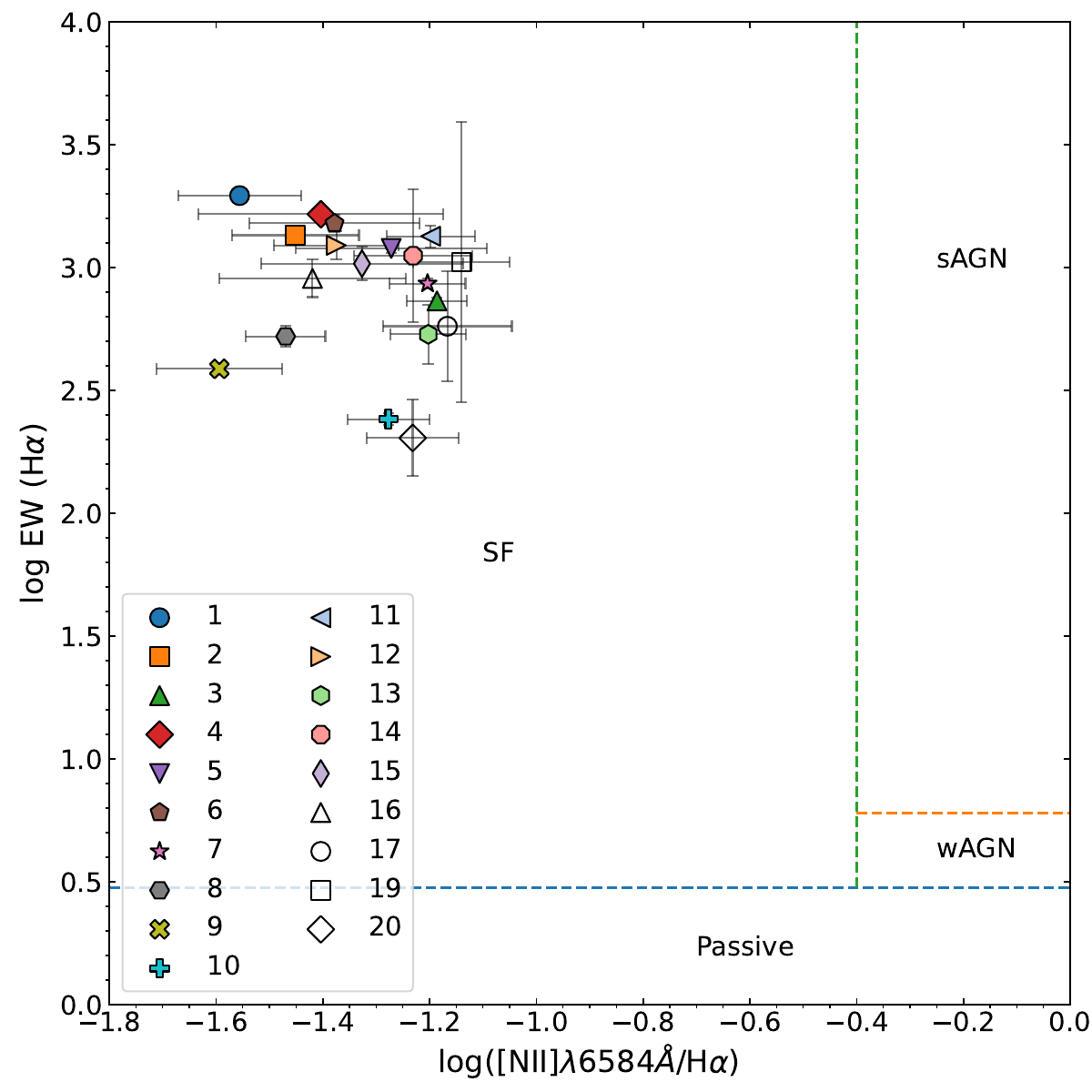}
\caption{Diagnostic WHAN diagram \citep{cidfernandes+2011} for the 19 star-forming regions observed in NGC 1313. The dashed green line separates AGNs and star formation, the dashed orange line separates strong and weak AGNs and the dashed blue line separates star-forming from passive galaxies.}
\label{fig:whan}
\end{figure}

We used diagnostic diagrams to determine the dominant ionization mechanism in the observed regions. Using the Baldwin–Phillips–Terlevich (BPT; \citealt{bpt1981}) and EW$_{H\alpha}$ versus N[\textsc{ii}]/H$\alpha$ (WHAN; \citealt{cidfernandes+2011}) diagrams, we find that all 19 regions correspond to star-forming regions, i.e., H\textsc{ii} regions, as shown in Figures \ref{fig:bpt_NII} and \ref{fig:whan}. Hence, the gas is ionized by UV emission from massive O- and B-type stars. One region (ID \#19) lies at the limit of the star-forming regime in the top panel in Figure \ref{fig:bpt_NII}; however, within the uncertainties, this source appears to be a star-forming region. In the bottom panel of Figure \ref{fig:bpt_NII}, region \#19 shows an atypical [S{\sc ii}]/H$\alpha$ ratio. This atypical ratio may be due to its spatial position between the two complex southwestern zones. Therefore, an additional classification scheme is necessary to elucidate its origin.

Figure \ref{fig:whan} presents the WHAN diagnostic diagram, which uses EW(H$\alpha$) instead of [O\textsc{iii}]/H$\beta$. All regions lie in the star-forming zone. These results confirm that region \#19 is an outlier in the previous diagnostic diagrams and is ionized by star formation. Furthermore, this confirms that the observed regions are H$\textsc{ii}$ regions.

\subsection{Electron densities}

We measured the [S\textsc{ii}] doublet ratio (RS2) in all 19 H\textsc{ii} regions, as we detected both emission lines in each case. However, six regions exhibit RS2 values above the upper theoretical limit of 1.43 \citep{OF2006}, indicating that their electron densities fall within the low-density limit. These regions are located close to the center (region \#3 and \#10) and in the southwestern area (region \#13, \#14, \#15, and \#19). For these regions, we assumed a value of $n_e<$10 cm$^{-3}$. In the remaining 13 regions, where the ratio lies within the valid diagnostic range (specifically, 1.28 $<$ RS2 $<$ 1.43), we estimated electron densities. We list the derived electron densities for all regions in the sixth column of Table \ref{tab:prop_fisicas}. The electron densities in NGC 1313 span the range $n_e$ = 10 to 142 cm$^{-3}$, with a median of $n_e$ = 71 cm$^{-3}$ (16--84th percentiles: 10--111 cm$^{-3}$), which is consistent with the low-density regime of H\textsc{ii} regions. In contrast to ultra-compact H$\textsc{ii}$ regions (with $n_e>10^4 - 10^5$ cm$^{-3}$), these are more diffuse regions, as suggested by \citet{copetti+2000}, who found low mean electron density values ranging from  $n_e\approx$20 to 140 cm$^{-3}$ in a sample of 15 galactic H\textsc{ii} regions using the RS2 indicator.  \citet{krabbe+2014} found that the expected values for $n_e$ in interacting systems were in the range $n_e$ = 24 to 532 cm$^{-3}$. However, their sample consisted of pairs of interacting galaxies and in some cases major mergers, whereas NGC 1313 does not belong to a galaxy pair. Furthermore, they studied mostly SA-type galaxies. However, their sample included one SB-type galaxy (AM-2058B, similar to NGC 1313), with a mean $n_e$ = 86 cm$^{-3}$, spanning the range $n_e$ = 42 to 184 cm$^{-3}$. These values are similar to those obtained in this work.

\newcommand{\ageasym}[3]{\ensuremath{#1^{+#3}_{-#2}}} 
\newcommand{\valerr}[2]{%
  \ifthenelse{\equal{#2}{} \OR \equal{#2}{-}}%
    {#1}%
    {#1\,$\pm$\,#2}%
}
\newcommand{\valerrr}[3]{%
  \ifthenelse{\equal{#3}{} \OR \equal{#3}{-}}%
    {#1}%
    {#1\,$\pm$\,#2\,($\pm$\,#3)}%
}
\begin{table*}\label{tab:prop_fisicas}
  \small
  \centering
  \caption{Deprojected distances, N2 calibrator, oxygen abundances, electron densities, and ages for the star-forming regions in NGC 1313.}  
  \label{tab:prop_fisicas}
  \hspace*{-30pt}
  \begin{tabular}{ccccccc}
    \hline
    & & & & & \\ 
    \multirow{2}{*}{ID} 
      & \multirow{2}{*}{r [kpc]$^a$} 
      & \multirow{2}{*}{N2} 
      & \multirow{2}{*}{12+log(O/H)$^b$$^\dagger$}  
      & \multirow{2}{*}{RS2$^c$} 
      & \multirow{2}{*}{$n_e$ [$\mathrm{cm}^{-3}$]$^d$}
      & \multirow{1}{*}{Ages $^e$}\\
    & & & & & & [$10^6$ yr]\\
    & & & & &\\ 
    \hhline{-------}
    & & & & &\\ 
    \begin{filecontents*}{tablas/new_unc_gradiente_scarano_ne_masa_edad_err_newnumber_ne10_menos.csv}
ID,O3N2,O3N2_ERR,12_LOG_OH_O3N2,12_LOG_OH_O3N2_ERR,12_LOG_OH_O3N2_TOTAL_ERR,N2,N2_ERR,12_LOG_OH_N2,12_LOG_OH_N2_ERR,12_LOG_OH_N2_TOTAL_ERR,r_kpc,RS2,RS2_err,n_e,mass,age,age_err_high,age_err_low,r_kpc_un
1,2.30,0.12,-,,,-1.55,0.11,8.02,0.07,0.18,1.98,1.30,0.03,121,5.75,2.73,0.02,0.02,0.11
2,2.15,0.12,-,,,-1.45,0.11,8.07,0.07,0.18,2.21,1.35,0.03,71,5.94,3.33,0.46,0.24,0.12
3,1.66,0.06,8.18,0.03,0.18,-1.18,0.05,8.20,0.05,0.17,0.91,1.57,0.10,$<$10,6.11,4.74,0.03,0.02,0.02
4,2.04,0.23,-,,,-1.40,0.22,8.10,0.11,0.20,1.63,1.31,0.03,110,5.77,2.87,0.04,0.03,0.04
5,1.69,0.18,8.17,0.05,0.19,-1.27,0.17,8.16,0.09,0.19,2.06,1.32,0.02,100,5.75,3.65,0.19,0.33,0.08
6,2.05,0.16,-,,,-1.37,0.15,8.11,0.09,0.18,4.01,1.32,0.03,100,5.50,2.99,0.23,0.11,0.23
7,1.63,0.07,8.18,0.03,0.18,-1.20,0.07,8.19,0.05,0.17,4.64,1.42,0.05,10,5.13,4.57,0.06,0.07,0.27
8,2.11,0.08,-,,,-1.46,0.07,8.06,0.06,0.17,2.73,1.42,0.03,10,5.57,4.93,0.03,0.03,0.14
9,1.83,0.12,-,,,-1.59,0.11,8.01,0.07,0.18,1.43,1.28,0.05,142,6.11,5.03,0.04,0.03,0.01
10,1.75,0.08,-,,,-1.27,0.07,8.15,0.05,0.17,1.17,1.69,0.13,$<$10,6.12,5.77,0.09,0.11,0.02
11,1.60,0.09,8.19,0.03,0.18,-1.19,0.08,8.19,0.05,0.17,5.02,1.41,0.04,18,5.27,3.40,0.30,0.35,0.16
12,1.87,0.12,-,,,-1.37,0.11,8.11,0.07,0.17,4.17,1.42,0.05,10,5.04,3.88,0.25,0.68,0.12
13,1.47,0.08,8.22,0.03,0.18,-1.20,0.07,8.19,0.05,0.17,4.35,1.56,0.07,$<$10,4.97,4.93,0.10,0.14,0.14
14,1.81,0.12,-,,,-1.23,0.11,8.17,0.06,0.17,3.47,1.48,0.08,$<$10,5.26,3.25,1.74,0.44,0.06
15,1.80,0.19,-,,,-1.32,0.18,8.13,0.10,0.19,7.11,1.46,0.04,$<$10,4.99,4.22,0.33,0.55,0.39
16,1.94,0.18,-,,,-1.41,0.17,8.09,0.09,0.18,7.05,1.36,0.07,62,4.94,4.50,0.23,0.35,0.38
17,1.53,0.13,8.21,0.03,0.18,-1.16,0.12,8.20,0.07,0.17,6.31,1.37,0.07,53,5.58,4.90,0.48,0.35,0.35
19,1.97,0.11,-,,,-1.14,0.09,8.22,0.06,0.17,4.97,1.44,0.15,$<$10,4.77,4.17,4.17,4.17,0.23
20,1.23,0.16,8.27,0.04,0.18,-1.23,0.08,8.17,0.06,0.17,6.20,1.31,0.14,110,4.58,6.02,0.76,0.61,0.33
\end{filecontents*}
\csvreader[
      late after line=\\, 
      late after last line=\\\hline
    ]{tablas/new_unc_gradiente_scarano_ne_masa_edad_err_newnumber_ne10_menos.csv}{}{
      \csvcoli & 
      \valerr{\csvcolxii}{\csvcolxx} & 
      \valerr{\csvcolvii}{\csvcolviii} & 
    \valerrr{\csvcolix}{\csvcolx}{\csvcolxi} &  
      \valerr{\csvcolxiii}{\csvcolxiv} & 
      \csvcolxv &
      \ageasym{\csvcolxvii}{\csvcolxix}{\csvcolxviii}}
  \end{tabular}\vspace{2mm}
  \begin{minipage}{\textwidth}
  \raggedright
  $^a$ Deprojected galactocentric distance of the star-forming regions from the center of NGC 1313, estimated using the method proposed by \citet{scarano+2008}.\\
  $^b$ Oxygen abundances estimated using the N2 method proposed by \citet{marino+2013}.\\
  $^c$ Line ratio [S\textsc{ii}]\ $\lambda$6716\AA/[S\textsc{ii}]\ $\lambda$6731\AA. \\
  $^d$ Electron densities obtained with the \textsc{temden} task of \textsc{IRAF}.\\
  $^e$ Age of the region estimated using the \textsc{Starburst99} model. Uncertainties correspond to the confidence interval derived from the H$\alpha$ equivalent width (EW) uncertainty.\\
  $^{\dagger}$ The uncertainty values are estimated from the uncertainties associated with the emission-line flux measurements. In parentheses we include the uncertainty associated with the dispersion of the calibrator (0.16 dex).
  \end{minipage}
\end{table*}

\subsection{Stellar masses}

Table \ref{tab:mass} lists the derived magnitudes and masses. The masses range from $10^{4.9}$ to $10^{6.3}$ M$_\odot$. This indicates that the regions host stellar populations spanning tens of thousands to a few million solar masses, comparable in order of magnitude to well-studied massive star-forming complexes such as 30 Doradus in the LMC (M$_\star\sim1.1\times10^{5}M_\odot$; \citealt{doran+2013}) and the young massive clusters recently studied with JWST in the barred spiral galaxy NGC 1365 ($ M_\star\gtrsim10^{6}M_\odot $; \citealt{whitmore+2023}). The stellar mass values are similar to those found by \citet{finn+2024a}, who note that NGC 1313 is particularly efficient at forming massive clusters (M$_\star>10^{5}$M$_\odot$) compared to NGC 7793, a spiral galaxy with similar physical properties. 

\begin{table*}
  \centering
  \small
  \caption{Apparent magnitudes in the $g$ and $i$ bands, absolute magnitudes in the $i$ band, and masses with their uncertainties for the H$\textsc{ii}$ regions in NGC 1313.}
  \label{tab:mass}
  \begin{tabular}{ccccc}
         \hline
         & & & & \\
         \multirow{2}{*}{ID} & \multirow{2}{*}{$m_g^{*a}$} & \multirow{2}{*}{$m_i^{*a}$} & \multirow{2}{*}{$M_i^{*b}$} &  \multirow{1}{*}{Mass$^{*c}$ }  \\ 
          & & & & log(M$_*)$[$M_{\odot}$] \\
          & & & & \\
         \hhline{-----}
         & & & &\\
    \csvreader[
      head to column names,
      late after line=\\,
      late after last line=\\\hline
    ]{\detokenize{tablas/mag_mass_age_err_phot.csv}}{}%
    {%
      \csvcoli 
      & \valerr{\csvcolvi}{\csvcolvii} 
      & \valerr{\csvcoliii}{\csvcoliv} 
      & \valerr{\csvcolviii}{\csvcolix} 
      & {\valerr{\csvcolx}{\csvcolxi}} 
    }
  \end{tabular}\vspace{2mm}
  \begin{minipage}{\textwidth}
  \raggedright
  
  $^a$ Apparent magnitude corrected for extinction \citep{calzetti+2000} in the $g$ and $i$ bands of the Legacy Survey.\\
  $^b$ Absolute magnitude in the $i$ band of the Legacy Survey.\\
  $^c$ Stellar mass.\\
  $^*$  The uncertainties associated with the apparent magnitudes are obtained from the Legacy weight image for each filter and aperture. The uncertainties for the stellar masses are then propagated.
  \end{minipage}
\end{table*}

\subsection{Ages from H$\alpha$ equivalent width}

We inspected the spectra of the star-forming sources in NGC 1313 and found that the stellar continuum is too faint to be detected (e.g., see the panel in Figure \ref{fig:ngc1313_slits_center}). In addition, we detect Wolf-Rayet Bumps in at least two regions (\#5 and \#7), which indicate a very young population with ages in the range of 3 to 5 Myr \citep{hadfield+2007}.  \cite{silva-villa+2012} suggested the presence of a burst in the southwestern region based on an analysis of star formation histories in the system. This evidence suggests that the regions formed in an instantaneous burst of star formation rather than through continuous star formation.

As mentioned in Section \ref{sec:ages}, we estimated ages using \textsc{SB99}. In practice, we used a simple stellar population, the Salpeter IMF and a metallicity of Z=0.004, which is consistent with the abundances derived from the N2 method (average value of 12+log(O/H)$\approx$8.13; see section \ref{sec:oxygen_results}). We list the estimated ages in the last column of Table \ref{tab:prop_fisicas}.

The ages obtained span $2.73$-$6.02$ Myr, with a mean of $\sim 4.2$ Myr, confirming a uniformly young population consistent with typical H\textsc{ii} regions powered by massive O-type stars. In detail, eight regions are younger than 4Myr, eight lie between 4--5Myr, and three are older than 5Myr. These timescales are consistent with the short lifetimes of O stars ($\sim$ 4Myr) and the expectation that H\textsc{ii} regions remain ionized for $\lesssim 10$ Myr \citep{OF2006, draine2011}.

\subsection{Oxygen abundances and gradient}\label{sec:oxygen_results}

The oxygen abundances considered in this work are those derived with the N2 calibrator, with all 19 regions falling within its valid range. Using the N2 calibrator, the oxygen abundances range from 12+log(O/H)=8.01$\pm$0.18 (region \#9, northeast arm) to 12+log(O/H)=8.22$\pm$0.17 (region \#19, southwest sector), with a mean of 12+log(O/H)=8.13$\pm$0.04. These values confirm that NGC 1313 is a low-metallicity system, with sub-solar abundances compared with 12+log(O/H)$_\odot$=8.69 \citep{allende-prieto+2001}.

The derived abundances are systematically lower than previously estimated with the direct method (T$_e$), such as 12+log(O/H)$\approx$8.26$\pm$0.07 from \citet{pagel+1980}, 12+log(O/H)$\approx$8.4$\pm$0.1 by \citet{walsh+1997}, and most recently 12+log(O/H)$\approx$8.26$\pm$0.06 from \citet{hadfield+2007}. This discrepancy arises from the different methods used to estimate abundances. In the aforementioned studies, the auroral line [O\textsc{iii}] $\lambda$4363\AA\ could be detected, while in this work we used semiempirical calibrators. However, our analysis uses more recent calibrations based on updated statistical fits. Furthermore, we covered more regions in the southwestern area than previous studies that estimated the abundances using the direct method.

\begin{figure}[h!]
    \centering
    \includegraphics[width=1.0\linewidth]{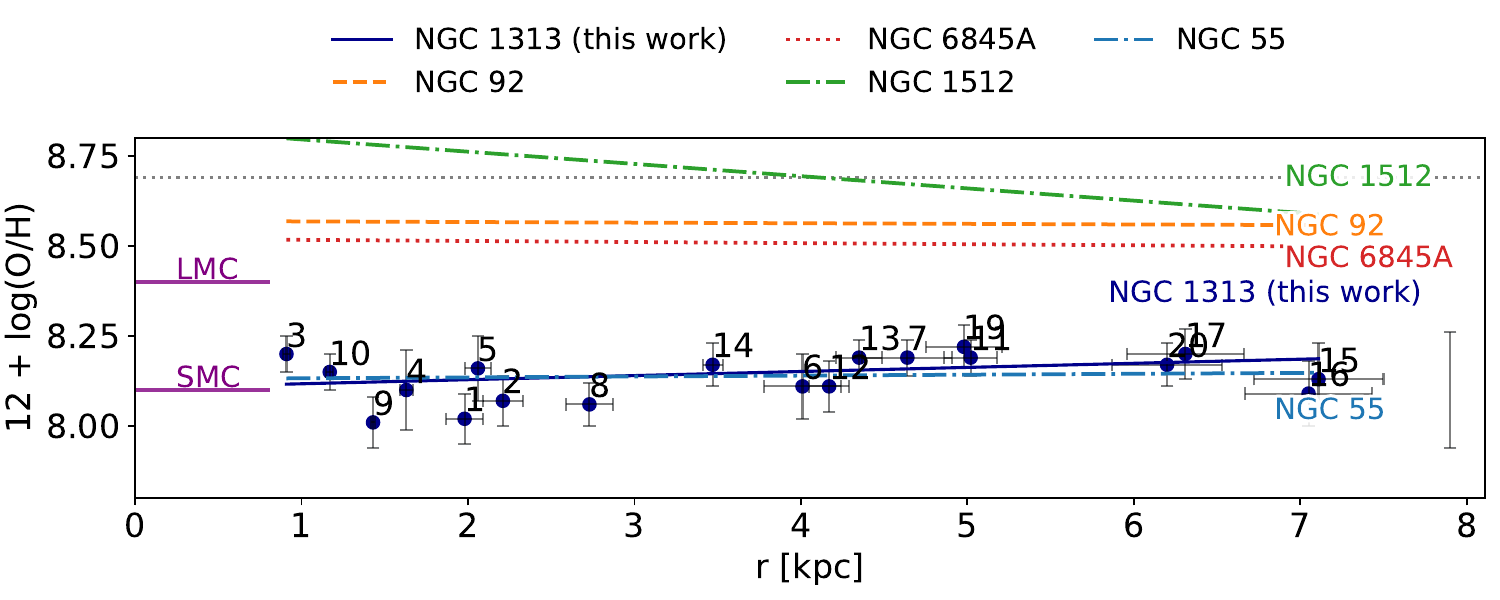}
    \caption{Radial abundance gradient in NGC 1313. The points represent the abundances derived using the N2 method for the 19 regions studied. The error bars for each star-forming region represent the uncertainties in the oxygen abundances, including the flux estimation errors and the uncertainties in the inclination and position angle. For reference, we show the calibrator dispersion (0.16 dex; \citealt{marino+2013}) as a vertical line. For comparison, we show the slopes of different interacting systems, such as NGC 92 \citep{torres-flores+2014}, NGC 6845 \citep{olave-rojas+2015}, NGC 1512 \citep{bresolin+2012}, and NGC 55 \citep{magrini+2017}. The average abundances of the LMC and SMC \citep{russell+1990} are also indicated for reference.}
    \label{fig:gradiente_comparacion}
\end{figure}

Fig. \ref{fig:gradiente_comparacion} presents the radial distribution of the oxygen abundance in NGC 1313, spanning deprojected distances of 0.91 to 7.11 kpc from the galactic center. We adjusted the linear fit for all regions using a least-squares method with \textsc{numpy.polyfit}, including uncertainties in abundance measurements, yielding a weakly inverted gradient with a slope of $\beta$ = 0.0114$\pm$0.0074 dex kpc$^{-1}$ and an intercept of {12+log(O/H) = 8.106$\pm$0.030}. In particular, we find the highest abundances in the southwestern regions, with region \#19 being the most metal-rich and located between the two southwestern structures. In contrast, we find the lowest abundances in regions \#1 and \#9, which are located in the inner parts of the northeastern and southwestern spiral arms, respectively. However, this ``inverted gradient'' is considered a marginal trend because, despite the positive slope, the uncertainties suggest a mostly flat radial chemical distribution rather than an inverted gradient. 

For comparison, Fig. \ref{fig:gradiente_comparacion} includes the chemical gradients for other interacting systems: NGC 55, a barred spiral galaxy with a stellar mass of 3.0$\times$10$^9$M$_\odot$ \citep{medoff+2025}, which has a very similar slope and the same average oxygen abundance (12+log(O/H)$\sim$8.13$\pm$0.18; \citealt{magrini+2017}). We also included three more massive interacting systems (on the order of $\approx$10$^{10}$M$_\odot$): NGC 1512, a barred spiral \citep{bresolin+2012}, and the regions observed in the tails of NGC 92 \citep{torres-flores+2014} and NGC 6845 \citep{olave-rojas+2015}. Consistent with the mass-metallicity ratio, these more massive systems exhibit higher oxygen abundances. However, the same flattening of the chemical gradient is still present. In addition, we plot the average oxygen abundance value estimated for the LMC and the Small Magellanic Cloud (SMC; \citealt{russell+1990}). The oxygen abundance of NGC 1313 is similar to the average value estimated for the SMC.

It is important to note that we calculated the uncertainties associated with the slope and the zero-point of the fits from the uncertainties in the inclination, position angle, and flux without adding the 0.16 dex scatter of the calibrator \cite{marino+2013}. This is because such a dispersion increases the uncertainty, making it difficult to distinguish between flat, positive, or negative gradients. Nevertheless, since we determined the oxygen abundance from the N2 index, our measurements are not absolute values and depend on the calibrator. Therefore, the 0.16 dex scatter becomes systematic without altering the general trends. Table \ref{tab:slopes} lists the slopes and zero-points for the different fits applied to the chemical distributions.

\begin{figure}[h!]
    \centering
    \includegraphics[width=\columnwidth]{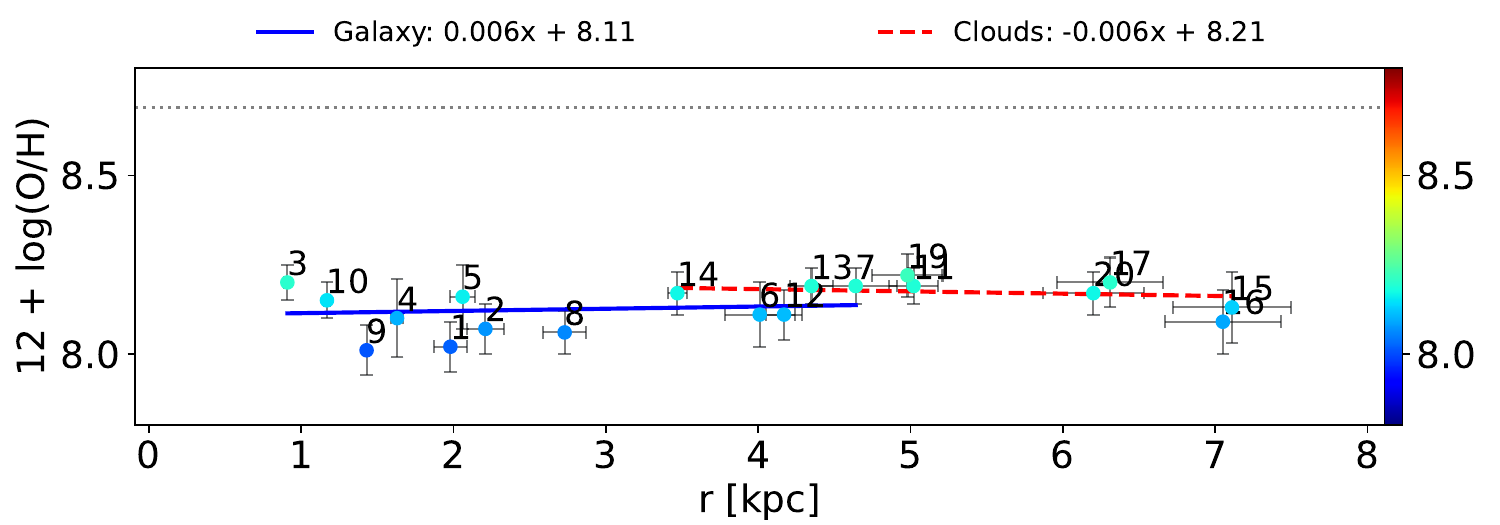}\\
    (a)
    \includegraphics[width=\columnwidth]{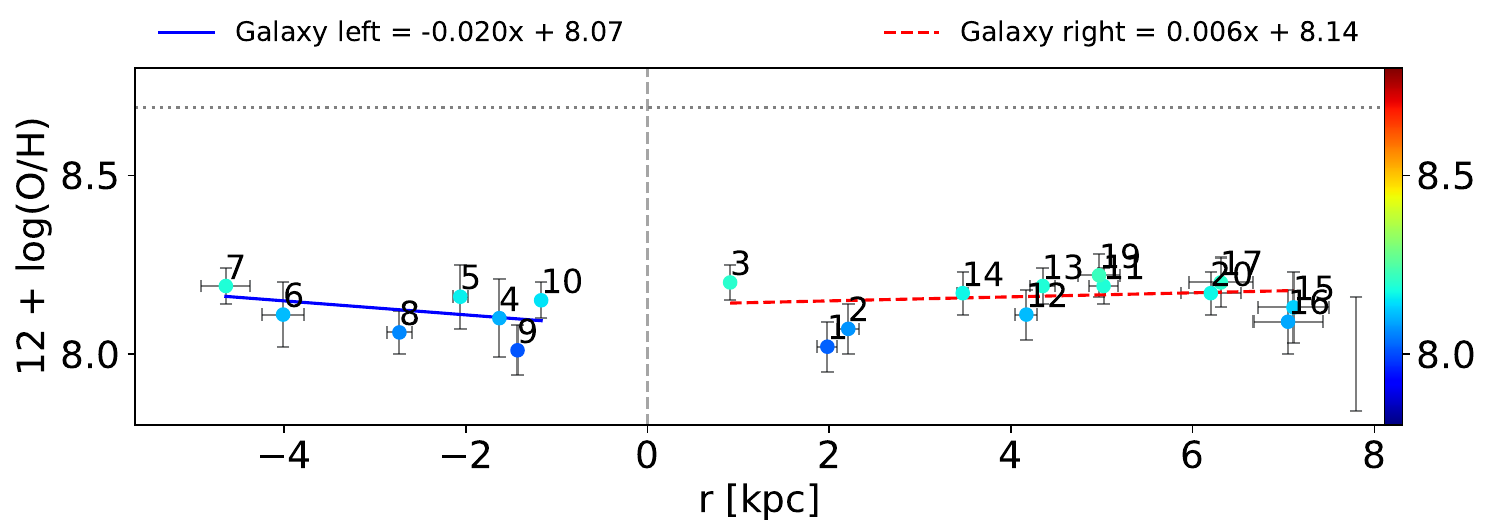}
    (b)
    \caption{Oxygen abundance gradients for the 19 star-forming regions using the N2 method from \citet{marino+2013}. The error bars for each star-forming region represent the uncertainties in the oxygen abundances, including the flux estimation errors and the uncertainties in the inclination and position angle. For reference, we show the calibrator dispersion (0.16 dex; \citealt{marino+2013}) as a vertical line. Panel (a): Radial gradient separating regions belonging to the main body of the galaxy (``Galaxy'') and the southwestern regions of the system (``Clouds''). Panel (b): Galactocentric gradient separating between regions northeast (``Galaxy left'') and southwest (``Galaxy right'') of the center of NGC 1313. For all gradients, the slope and intercept of the fits are indicated at the top.}
    \label{fig:gradients}
\end{figure}

The marginal positive gradient in NGC 1313 is mainly driven by the southwestern area, where regions at 5-8 kpc exhibit higher abundances (12+log(O/H)$\sim$8.10 to 8.22, with a mean of 12+log(O/H)$\sim$8.16) than those in the main body at 1.5-5 kpc (12+log(O/H)$\sim$8.01 to 8.20, mean 12+log(O/H)$\sim$8.11). To further investigate this trend, we performed two different fits: one for the main galaxy (regions \#1 to \#10) and another for the southwestern sector (regions \#11 to \#20), as shown in panel (a) of Fig. \ref{fig:gradients}. The regions in the main body of the galaxy maintain a flat gradient with a slight positive slope ($\beta$ = 0.0062 $\pm$ 0.0183 dex kpc$^{-1}$), whereas the clouds to the southwest now show a slightly negative slope ($\beta$ = -0.0064 $\pm$ 0.0121 dex kpc$^{-1}$). Again, considering the uncertainties, the trend is consistent with a mostly flat gradient. This trend is in agreement with previous studies suggesting that NGC 1313 lacks a pronounced radial gradient. \citet{walsh+1997} found that the global oxygen abundance distribution is essentially flat across the disk and pointed out that NGC 1313 is one of the most massive barred galaxies that does not exhibit a radial gradient. Although the oxygen abundance values in this work are lower (12+log(O/H)$\sim$8.0 to 8.2 compared to 12+log(O/H)$\sim$8.4 obtained by \citet{walsh+1997}, possibly due to the different methods used), we still observe a flat gradient. Specifically, we detect a slight increase in abundance in the southwestern region of the system.

To analyze how the gas distribution behaves in each arm of NGC 1313, we created a galactocentric gradient which divides the galaxy into the northeastern arm and the southwestern region (southwestern arm and structures), as shown in panel (b) of Figure \ref{fig:gradients}. This gradient shows that the outermost H\textsc{ii} regions of both the northeastern and southwestern sides of the galaxy are more metallic than the center. In particular, for the northeast arm we find a slope of $\beta$=-0.0196$\pm$0.0176 with an intercept of 12+log(O/H)=8.070$\pm$0.052, while for the southwestern region we find a slope of $\beta$=0.0057$\pm$0.0094 with an intercept at 12+log(O/H)=8.137$\pm$0.043.

\

\begin{table}
  \centering
  \caption{Slopes and intercepts calculated for different fits to the oxygen abundance gradient in NGC 1313.}
  \label{tab:slopes}
    \resizebox{\columnwidth}{!}{%
  \begin{tabular}{cccc}
    \hline
    & & & \\
    Gradient & Fit & Slope$^a$ & Intercept$^a$ \\
    & & & \\
    \hhline{----}
    & & & \\
    \multirow[c]{3}{*}{Radial} &
      Galaxy and clouds & 0.0114 $\pm$ 0.0074 & 8.106 $\pm$ 0.030 \\* 
    & Only galaxy & 0.0062 $\pm$ 0.0183  & 8.108 $\pm$ 0.048 \\*   
    & Only southwest clouds & -0.0064 $\pm$ 0.0121 & 8.207 $\pm$ 0.063 \\*
    & & & \\
    \hline
    & & & \\
    \multirow[c]{2}{*}{Galactocentric} &
      Northeast arm & -0.0196 $\pm$ 0.0176 & 8.070 $\pm$ 0.052 \\*   
    & Southwest arm and clouds & 0.0057 $\pm$ 0.0094 & 8.137 $\pm$ 0.043 \\*   
    \hline 
  \end{tabular}\vspace{2mm}
  }
    \begin{minipage}{\columnwidth}
  \raggedright
    $^a$ The uncertainties are calculated from the uncertainties in the flux estimations, inclination, and position angle. 
    \end{minipage}
\end{table}

\begin{figure*}
\centering 
    \includegraphics[width=0.6\textwidth]{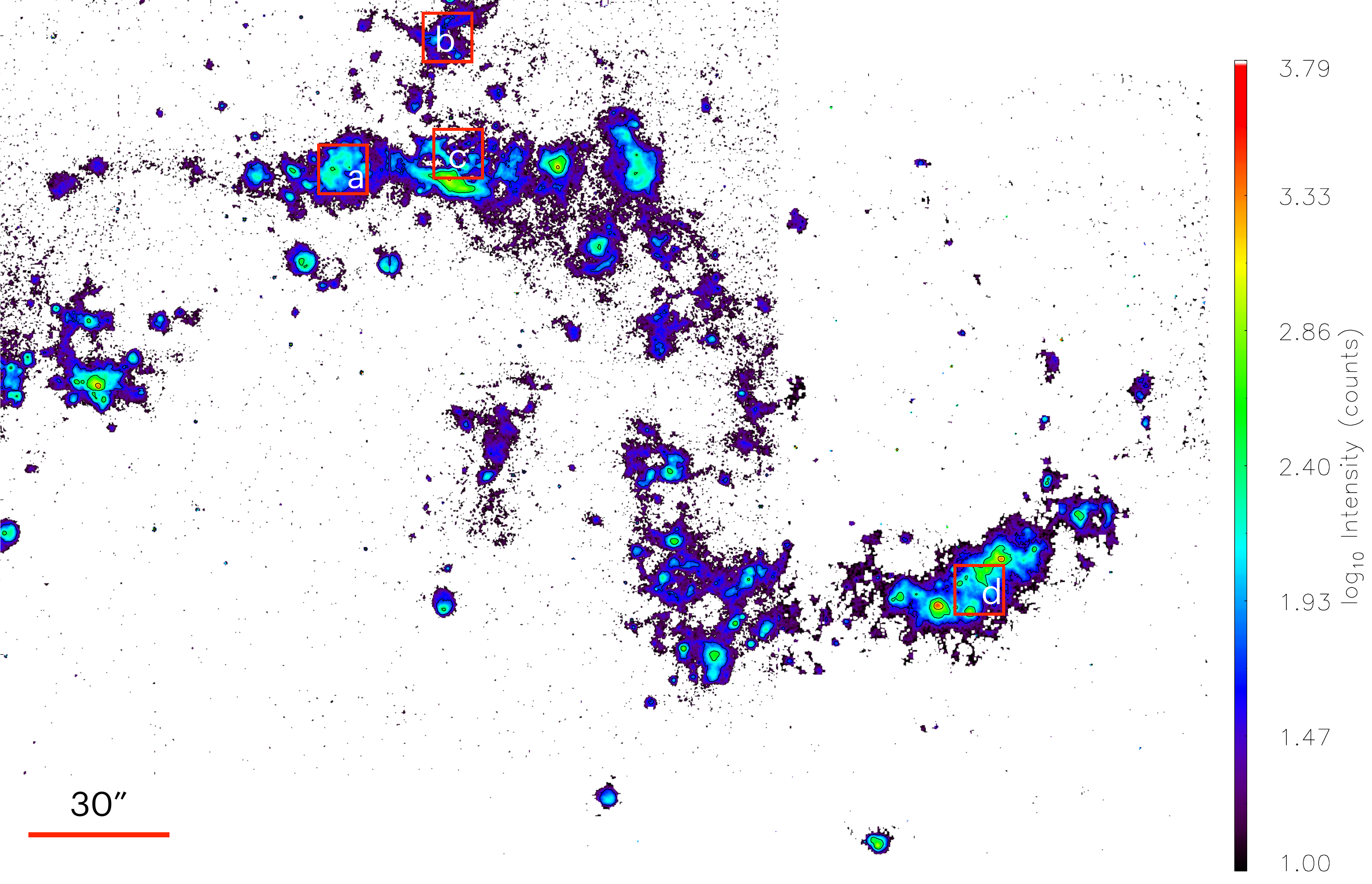}\\
    \includegraphics[width=0.6\textwidth]{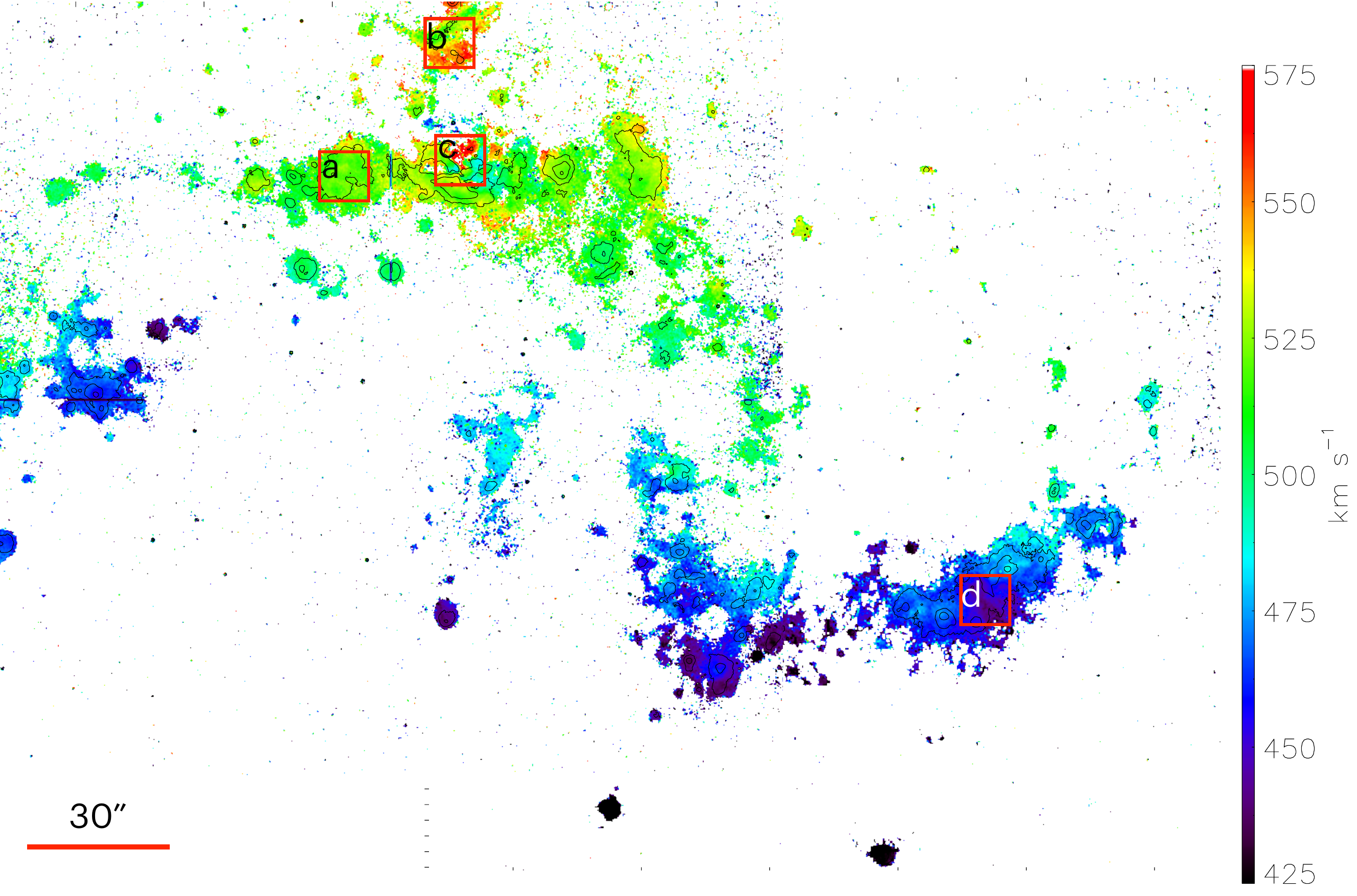}\\
    \includegraphics[width=0.8\textwidth]{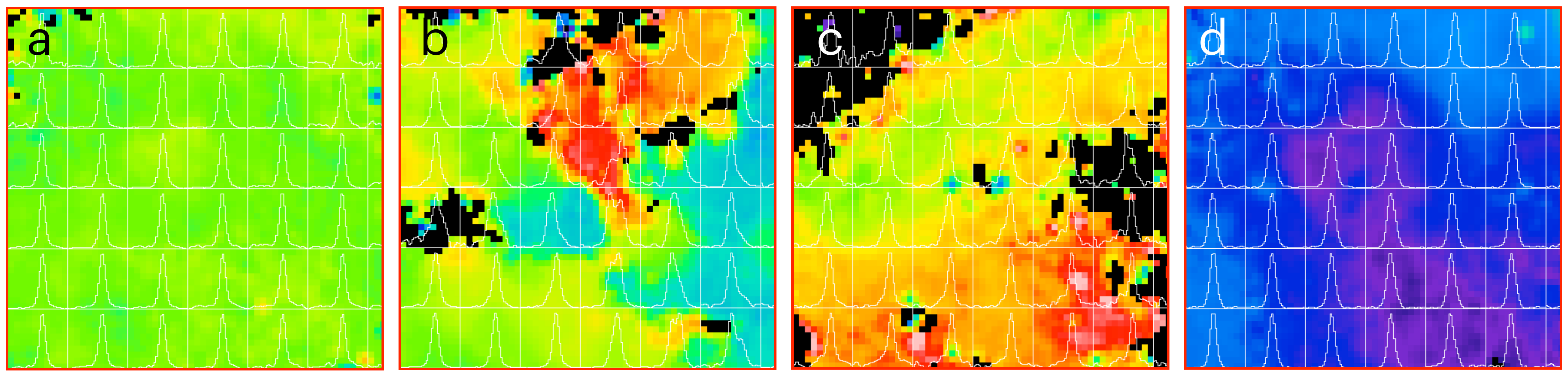}
    \caption{Top panel: H$\alpha$ image of NGC 1313 derived from the SOAR Adaptive Module-Fabry-Perot (SAM-FP) data cube. Middle panel: Velocity field for NGC 1313. In both panels, slack contours represent the H$\alpha$ emission and selected regions are highlighted with red boxes. Bottom panel: Zoom-in of regions showing the complex kinematics of NGC 1313. Each panel shows the H$\alpha$ profiles extracted from 10$\times$10 pixel boxes in the data cube.}
    \label{fig:kinematic}
\end{figure*}

\subsection{H$\alpha$ distribution and kinematic behavior}

In the top panel of Figure \ref{fig:kinematic} we show the H$\alpha$ monochromatic image of NGC 1313, which corresponds to a mosaic of two different pointings. In this figure we identify the northeastern and the southwestern spiral arms. Both are clearly identified in the H$\alpha$ image, which traces star formation on this system. We observe patchy emission across the arms. We do not identify strong emission in the central region, which indicates the presence of a more evolved population that cannot ionize gas. In addition, we observe different H\textsc{ii} regions outside the spiral arms, which may suggest previous interaction events. 

The middle panel of Figure \ref{fig:kinematic} shows the velocity field of NGC 1313. In general, there is a clear velocity gradient in this system, which indicates that the galaxy is gravitationally supported by rotation. However, on small scales there are several deviations from a purely rotating disk. To highlight this point, the bottom panel of Figure \ref{fig:kinematic} shows four panels illustrating the different complexity of this system, where white profiles represent the H$\alpha$ emission extracted from 10$\times$10 pixel boxes. The spatial location of these panels is identified by red boxes in the middle and top panels of Figure \ref{fig:kinematic}.

In panel a, the velocity field appears remarkably homogeneous, with no significant gradients or localized perturbations in the radial velocity. The H$\alpha$ line profiles are predominantly symmetric and well described by a single component, indicating that the ionized gas is dominated by ordered motions, likely tracing the underlying disk rotation. This suggests a dynamically relaxed region with no clear signatures of recent disturbances or strong local feedback. In contrast, panel b exhibits markedly complex kinematics. The radial velocity shows strong spatial variations on small physical scales, reaching amplitudes of up to $\sim$60 km s$^{-1}$. The H$\alpha$ profiles in this region, particularly near the central area, frequently display multiple components and asymmetric shapes. Such features are indicative of kinematically distinct gas components along the line of sight, which may arise from expanding shells, superbubbles driven by stellar feedback, or localized gas flows. Alternatively, these signatures may reflect noncircular motions induced by a past or ongoing interaction event, as observed in systems such as HCG 31 (e.g., \citealt{Amram2007}). Panel c also reveals disturbed kinematics, although somewhat less extreme than in panel b. The velocity field is irregular, with variations of up to $\sim$50 km s$^{-1}$ over spatial scales of $\sim$200 pc. The presence of perturbed and occasionally asymmetric H$\alpha$ profiles further supports the existence of localized dynamical disturbances. These may be associated with smaller-scale feedback processes (e.g., stellar winds or supernova-driven turbulence) or could represent residual signatures of larger-scale perturbations affecting the northern region of the galaxy. Panel d shows a region located in the western spiral arm of NGC 1313, where the velocity field is smooth and coherent, consistent with large-scale rotation. Although mild velocity gradients are present, they are not accompanied by distortions in the H$\alpha$ line profiles, which remain largely symmetric and single-peaked. This indicates that the gas kinematics in this region is dominated by regular disk dynamics, with no strong evidence for perturbations or multicomponent structures. Taken together, the kinematic analysis reveals a clear dichotomy between different regions of the galaxy. Regions a and d show a smooth behavior consistent with the large-scale pattern. Regions b and c show strong deviations from ordered rotation, including velocity gradients, asymmetries, and multicomponent profiles, all of which point to a dynamically disturbed gas. These features are consistent with the presence of noncircular motions, such as inflows, outflows, or tidal perturbations.

This qualitative and spatially resolved evidence supports a scenario in which NGC 1313 has undergone a past interaction or minor merger. Such events are known to induce gas flows, enhance turbulence, and produce complex velocity fields, particularly in localized regions. The concentration of these signatures in the northern part of the galaxy suggests that this area has been more strongly affected, possibly marking the most significant dynamical impact. In this context, previous interactions are expected to significantly affect the chemical distribution of the interstellar medium, as gas mixing acts as an efficient mechanism to redistribute metals throughout the galaxy (e.g., \citealt{rupke+2010a}). Such processes can dilute central metallicity enhancements and flatten radial abundance gradients by transporting metal-poor gas inward and metal-rich gas outward. From an observational perspective, interacting systems that exhibit disturbed kinematics often present shallow or flat metallicity gradients, consistent with the presence of large-scale gas flows (see, e.g., \citealt{AlfaroCuello15}). In this framework, the perturbed kinematics observed in NGC 1313 likely play a key role in shaping its chemical structure, as illustrated in Figure \ref{fig:gradients}. This is supported by the detection of multiple-component H$\alpha$ profiles, which provide strong evidence for the coexistence of kinematically distinct gas phases along the line of sight, a clear signature of ongoing gas mixing. These features support a scenario in which dynamical disturbance, such as inflows, outflows, or tidally induced motions, enhance the redistribution of metals, thereby contributing to the observed oxygen abundance pattern and its spatial variations.

\section{Discussion}\label{sec:discussion}

\subsection{Main physical properties of NGC 1313}

NGC 1313 is a late-type galaxy with a peculiar morphology, generally classified as a barred Magellanic spiral (SB(s)d). Due to its mass and luminosity, it lies at the boundary between dwarf and irregular galaxies and more developed disk spirals. Indeed, it has been classified as being in a ``transition'' between a barred Magellanic spiral (SBm) and an SBc \citep{walsh+1997}. Furthermore, it has been compared to the LMC, since both display a prominent bar and an irregular morphology. NGC 1313 is slightly less massive (2.6 $\times 10^9$ M$_\odot$; \citealt{calzetti+2015}) and, as confirmed by this work, has lower oxygen abundances (12+log(O/H)$\sim$8.13$\pm$0.04) than the LMC (12+log(O/H)$\sim$8.37; \citealt{russell+1990}). Notably, the values obtained are more similar to the abundance of the SMC (12+log(O/H)$\approx$8.13$\pm$0.1; \citealt{russell+1990}). 

All 19 observed regions in this study are confirmed as H$\textsc{ii}$ regions, whose dominant ionization source is clearly photoionization of massive stars, indicative of recent star formation. Region \#19, which appears as an outlier in the diagnostic diagrams, lies between the two molecular clouds in the southwestern area. Its atypical emission-line ratios may be enhanced by local conditions, possibly associated with the suggested past minor-merger interaction in the southwestern region of the system \citep{peters+1994}. It may also be associated with the nearby supernova SN 1978K and the presence of shocks reported by \citet{ryder+1993}. We cannot discard the possibility that shocks contribute to its ionization.

The derived electron densities indicate that most regions are in a low-density regime, with a median of $n_e\sim71$ cm$^{-3}$, suggesting the presence of diffuse nebular emission, specifically in the southwest structures. However, nine of the 19 regions fall within the $n_e$ range displayed by interacting galaxies \citep{krabbe+2014}, spanning  $n_e\sim$ 52 to 142 cm$^{-3}$. The densities obtained are generally lower than those reported by \cite{krabbe+2014}. One possible explanation is that NGC 1313 may have experienced a minor merger \citep{peters+1994, silva-villa+2012} rather than a major one, so its H\textsc{ii} regions do not reach such extreme electron densities. The observed median of $n_e\sim$71 cm$^{-3}$ is consistent with values reported for other H\textsc{ii} regions, such as N77A with $n_e\sim43$ cm$^{-3}$ and N456 with $n_e\sim114$ cm$^{-3}$, both in the SMC \citep{jin+2023}. Low densities are commonly found in H\textsc{ii} regions of interacting systems. For example, \citet{munoz-elgueta+2018} also found several regions with low values of $n_e$ ($<$10 to 20 cm$^{-3}$) in the interacting NGC 4656 system, which is also a nearby barred spiral galaxy with a dynamical mass of 6.8 $\times$ 10$^9$M$_\odot$ \citep{munoz-elgueta+2018}.

Regarding the spatial distribution, the southwestern regions tend to be less massive and less dense, but slightly older and more metal-rich. In contrast, the regions in the main body of the galaxy are generally the most massive, but we find no clear trend among age, density, or oxygen abundance.

\subsection{A young and massive stellar population}

The ages and stellar masses obtained for the H$\textsc{ii}$ regions in NGC 1313 confirm a young and massive population. This is expected since we selected these regions by their H$\alpha$ emission. Furthermore, considering that the galaxy has low metallicity and an intense H$\alpha$ emission, we also expected to find a young stellar population dominated by massive O- and B-type stars, which emit strongly in the UV. This implies recent and even ongoing star formation.

Our analysis shows that NGC 1313 hosts young ($<$ 6 Myr) and moderately massive ($\sim$ 10$^{4.9}$ to 10$^{6.3}$ M$_\odot$) H\textsc{ii} regions. This finding is in agreement with the work of \citet{finn+2024a}, who find that NGC 1313 is very efficient in forming young ($<$ 10 Myr) and massive ($>$10$^4$M$\odot$) star clusters, compared to the similar spiral NGC 7793. They attribute this efficiency to strong spiral density wave perturbations in NGC 1313, which lead to more extreme molecular cloud properties. The young ages we measure are also consistent with those found for star-forming regions in other interacting systems, such as the tidal tail of NGC 6845 reported by \citet{olave-rojas+2015}, who found ages ranging between 2-7 Myr.

Further evidence for a predominantly young stellar population is provided by \citet{hannon+2019}, who used photometric data to study $\sim$700 young clusters ($<$10 Myr) in NGC 1313 and in two other similar spirals (NGC 7793 and NGC 4395). They established a clear link between H$\alpha$ morphology and age: clusters with concentrated H$\alpha$ emission have an average age of $\sim$3 Myr, while those partially exposed by gas bubbles have ages close to $\sim$4 Myr. Finally, those largely devoid of ionized gas are older than 5 Myr. The regions observed in this work correspond to the younger categories in this evolutionary scenario, with the clusters still embedded in ionized gas.

Finally, the presence of the blue and red Wolf-Rayet bumps is also evidence for a young stellar population, specifically in regions \#5 and \#7. This provides an upper limit on the ages (3-5 Myr; \citealt{hadfield+2007}) and indicates that these regions host a massive stellar population, with O-type stars likely having masses $>$25M$_\odot$ \citep{gray+2009}. In particular, the ages of regions \#5 and \#7 are $\approx$ 3.65 and 4.57 Myr, respectively. This suggests that these regions were formed in situ.

\subsection{Flat and slightly inverted gradient: Metal mixing in the radial distribution}

The oxygen abundance gradient suggests the presence of a mixture of metals that causes flattening and even slight inversion of the gradient. Regardless of whether the N2 calibrator dispersion (0.16 dex) is included, a predominantly flat behavior is observed ($\beta$ = 0.0114 $\pm$ 0.0074 dex kpc$^{-1}$). This is expected based on several studies of interacting galaxies (\citealt{rupke+2010a}, \citealt{kewley+2010}, \citealt{perez+2011}, \citealt{rosa+2014}). For example, several tidal tails produced by galaxy-galaxy interactions display flat gradients, where star-forming regions trace these gradients, as in NGC 92 ($\beta$ = -0.017$\pm$0.0021; \citealt{torres-flores+2014}) and NGC 6845 ($\beta$ = 0.002$\pm$0.004; \citealt{olave-rojas+2015}). Subsequently, \citet{buzzo+2021} find flat and even inverted gradients when analyzing the NGC 1487 system, which is a merger of dwarf galaxies. Therefore, the results presented in this work are in agreement with the literature, which reports flat gradients in interacting galaxies. The aforementioned authors generally associate the flattening of the metal distribution with gas mixing through flows caused by gravitational interaction, which distribute the gas and dilute the central abundance.

However, it is important to emphasize that the flattening of the gradient may be triggered by gas flows rather than other factors, such as star formation. To this end, we used equation (1) from \citet{bresolin+2012}, which has been used in similar cases such as \citet{torres-flores+2014} and \citet{olave-rojas+2015}, to rule out the possibility that the star formation produced in situ by the interaction increased the abundance in the southwest region. Since we focused on the southwest part of the system, we adopted 12+log(O/H)=8.16 as the average abundance of regions \#11 to \#20. We calculated the star formation density using far-ultraviolet (FUV) images from the Galaxy Evolution Explorer (GALEX) through aperture photometry covering the regions of interest (using \textsc{photutils} from \textsc{python}). After correcting for extinction using the law of \citet{calzetti+2000} and calculating the star formation rate using equation (4) from \citet{iglesias-paramo2006}, we obtain $\Sigma_{SFR}=5.51\times10^{-6}$$M_\odot\ yr^{-1}\ kpc^{-2}$. For the neutral gas density, we assumed a lower bound of 10 $M_\odot\ pc^{-2}$, as proposed by Fig. 8 from \citet{suzuki+2013}, to avoid overestimation. \citet{silva-villa+2012} reported that 237 Myr are required to complete half an orbit around NGC 1313, which we adopted this as a maximum timescale. With these values, we obtain $y\sim3$ for an increase of 0.1 dex, which is much higher than the theoretical values ($y\sim0.01$ from \citealt{maeder1992} or $y\sim0.049$ from \citealt{lopez-sanchez+2015}). If we consider the $\Sigma_{SFR}$ observed with GALEX and the highest yield of 0.049, the increase in abundance is only 0.0016 dex, instead of the $\sim$0.1 dex we observed in the gradient. Similarly, reversing the calculation, increasing the abundance by 0.1 dex would require $\Sigma_{SFR}=3.8\times10^{-4}$$M_\odot\ yr^{-1}\ kpc^{-2}$, which is two orders of magnitude higher than our estimated value of $\Sigma_{SFR}=5.51\times10^{-6}$$M_\odot\ yr^{-1}\ kpc^{-2}$. It should be noted that the assumed $\Sigma_{HI}$ value may have been calculated for a larger region than the one of interest. For this reason, we repeated the calculation assuming half this value ($\Sigma_{HI}$=5 $M_\odot\ pc^{-2}$). Even with this calculation, we still need a $\Sigma_{SFR}=1.9\times10^{-4}$$M_\odot\ yr^{-1}\ kpc^{-2}$ to increase the abundance by 0.1 dex, obtaining a yield of 1.7. Conversely, with the calculated $\Sigma_{SFR}$ from GALEX, the abundance only increases by 0.003 dex. Consequently, we rule out the possibility that in situ star formation induced by the interaction itself could raise the oxygen abundance gradient, suggesting that gas flows are the most likely scenario for the flattening of the gradient.

\citet{walsh+1997} conducted a crucial study of the chemical analysis of NGC 1313 and noted that low-mass irregular galaxies such as the Magellanic Clouds or NGC 4395 exhibit flat gradients, while more massive late-type or Sc galaxies such as M33, NGC 300, or NGC 7793 develop typical negative gradients with slopes of $\beta$ = -0.08$\pm$0.02 dex kpc${^{-1}}$. Based on this, the authors determined that NGC 1313 was in an intermediate state, as it possessed a bar but not a prominent bulge, and given its flat gradient, they defined it as one of the most massive barred galaxies that does not exhibit a steep gradient. Similarly, the results obtained in this work confirm the homogeneous distribution of abundances, even when separated into different fits.

In contrast, a slight positive gradient trend has also been found in the Large Magellanic Cloud, where \cite{toribio_san_cipriano+2017} found a slope of $\beta$ = 0.05$\pm$0.05 using the collision-excited lines method and a slope of $\beta$ = 0.04$\pm$0.07 using the recombination lines method. Although these results differ in slope by $\approx$0.03 dex from those obtained in this work, they provide evidence that systems similar to NGC 1313 (such as the LMC) can also exhibit this inversion.

\citet{pan+2025} recently studied the evolution of the oxygen abundance gradient using Integral Field Unit (IFU) data from the Sloan Digital Sky Survey IV Mapping Nearby Galaxies at Apache Point Observatory (SDSS-IV MaNGA), analyzing galaxy pairs and interacting galaxies across different stages of the merger, from the first encounter to the final coalescence. With these separated states, the authors find that the gradient tends to flatten slightly just after the first pericenter passage, likely due to radial gas mixing. However, this flattening is not consistent in all cases and varies depending on the intensity of the interaction and star formation activity. The authors describe how, as the interaction progresses through different states, metallicity enrichment or dilution varies, resulting in gradients that can either flatten or become steeper than the initial gradient. Taking this into account, the flattening of the oxygen gradient observed in NGC 1313 may be associated with the first pericenter passage, i.e., we may be observing the gradient shortly after the first interaction proposed for NGC 1313.

\subsubsection{Past interaction with a satellite}

Although NGC 1313 has no neighboring galaxy with which it might show signs of  interaction, it shows a disturbance in its morphology and alterations in its physical and kinematic properties that could be due to past interaction. In particular, the southwest region stands out for its peculiarities. \citet{peters+1994} observed neutral hydrogen in NGC 1313 and found an extended loop of neutral hydrogen in the southwestern region. They subsequently found a kinematic perturbation when analyzing the velocity of the neutral gas. Based on this, they proposed that NGC 1313 interacted with a satellite galaxy (with a mass ratio of 1:10), which was destroyed and left this trace in the southwest region.

This scenario of a minor interaction was later revisited by \citet{silva-villa+2012}, who found a burst of star formation in the southwestern region triggered about 100 million years ago. They also emphasized that it was a minor burst, since they observed no evidence of interaction in other areas of the system, suggesting that the companion was a low-mass galaxy.

We observe a slight overabundance of oxygen in the southwestern region. This enrichment can be interpreted as a consequence of the recent starburst, where the intense formation of massive stars in that area tens or even 100 million years ago could have produced supernovae and, with them, associated stellar winds that may have enriched the interstellar medium with elements such as oxygen. In fact, there is a known Type II supernova (SNII) located in the southwestern region of NGC 1313, studied by \citet{ryder+1993} and designated SN 1978K, which has emitted intensely in X-rays, as well as in radio and optical spectra. This supernova may be an example of the intense events that occurred during that star formation episode, contributing chemical elements to the medium.

Now, the key question is how this supposed interaction influences the abundance gradient estimated in this work. Given that the gas in the southwestern region likely includes a mix of material native to NGC 1313 and that contributed by the satellite, the oxygen abundance in that region might be different. For example, if the satellite were a more metal-poor dwarf galaxy, a merger between the galaxies would tend to dilute the overall abundance by introducing low-metallicity gas. However, this work shows that the southwestern region is not poorer but rather more enriched in oxygen than the rest of the regions in the galaxy arms. This raises the issue of what is expected based on a smaller merger and its influence on the gradient. \citet{bustamante+2018} analyzed AURIGA cosmological simulations of wet mergers, considering progenitor galaxies with masses in the range 5 $\times$ 10$^9$ M$_\odot$ $<$ M$_\star$ $<$ 2 $\times$ 10$^{11}$ M$_\odot$, to study both minor mergers (mass ratios between 1:10 and 1:3) and major mergers (mass ratios between 1:3 and 1:1). The authors find that, while major mergers are known to strongly alter the radial distribution by diluting the metallicity, minor mergers also present a dilution in metallicity but with a significantly smaller amplitude. These results may explain why we observe a slightly flat gradient. In line with \citet{pan+2025}, only a major merger could invert the gradient or make it negative, as before the first pericenter. Thus, since NGC 1313 is  considered a possible minor merger by \citet{peters+1994}, the flattened gradient without a pronounced inversion is consistent with this interpretation.

It is important to note that the GMOS-S spectroscopic data cannot confirm the presence of an interaction. However, the results obtained suggest a slightly different behavior in the southwestern regions of the system. Considering this, together with information in the literature on NGC 1313, we suggest an evolutionary scenario in which this system underwent a minor interaction with a satellite that was cannibalized by NGC 1313. This triggered bursts of star formation in the southwestern region, where we observe low electron densities, SN 1978K, neutral hydrogen supershells, and young, massive regions. However, the flat gradient may be the result of gas mixing caused by gas flows. In this regard, we suggest that NGC 1313 likely experiences outflows driven by stellar winds, which enrich the medium by transporting elements from the central region. However, as mentioned in the simulation works (\citealt{rupke+2010a, perez+2011}), inflows of less enriched gas may also dilute the central abundance. In addition to this, we add the influence of the bar on the transport of gas toward the central regions of the Galaxy. For example, minor mergers alter gradients less dramatically than major mergers. As a result, one would expect only the southwestern region of the system to show this flattening, whereas the northeast arm (supposedly undisturbed) would not be affected. However, in this work, we also observe that the gradient in the northeast arm indicates flattening.

To complement and support the chemical analysis, the kinematic features reveal clear signatures of past interactions in NGC 1313. The northern region of this system reveals multiple H$\alpha$ profiles and noncircular motions, which suggest previous interactions. Conversely, the southwest spiral arm does not show strong evidence of kinematic anomalies. However, we note that our H$\alpha$ SAM-FP data do not cover southern star-forming complexes (as shown in Figure 1); therefore, we can only refer to the kinematics of the main body of this system. Overall, our kinematic findings are consistent with the observed oxygen abundance distribution, where a flatter chemical abundance can be explained by the gas mixing process due to previous interactions events. A spatially resolved spectroscopic analysis would be valuable for studying the correlation between the kinematics and the metal distribution in more detail.

\subsubsection{Influence of the bar}

The bar may also play a role in the mixing of chemical elements from one arm to the other, altering the chemical composition of the galaxy center. However, given the properties and regions observed in this work, it is not possible to determine the influence of each phenomenon on the flattening of the gradient. For that reason, we suggest that the observed flattening results from the influence of the bar, which drives radial gas flows that mix material throughout the galactic disk.

In NGC 1313, the bar is a prominent structure, so it is worth discussing whether it plays a role in gas mixing. This is because its role in gas dynamics is known, influencing gas flows towards the galactic center. \citet{walsh+1997} noted that barred galaxies have flatter radial abundance distributions compared to normal disk-type galaxies, regardless of their mass. Furthermore, they highlighted that a bar can be responsible for the transfer of angular momentum through radial gas flows into and out of the disk. Therefore, the bar becomes an effective pathway for radially homogenizing the abundance distribution \citep{friedli+1994}. \citet{walsh+1997} also noted that as the bar ages, the dilution of abundances by radial mixing decreases the slope; that is, it flattens the gradient. However, when observing regions in the bar of NGC 1313, they noted that it is metal-rich and hence young. One would expect a pronounced gradient, at least in the bar region, but they did not observe this in NGC 1313. Consequently, the authors suggest that the presence of the bar in galaxies such as NGC 1313 and the LMC results in the absence of a gradient. 

However, studies such as \citet{chen+2023}, which analyze oxygen abundance gradients in barred galaxies, find that the gradient break and flattening occur in regions farther from the center, suggesting that the flattening may be due to activity in the arms rather than in the bar. Similarly, simulations (e.g., \citealt{grand+2012, grand+2016}) show that spiral arms can induce significant radial migration and gas mixing, leading to flattened gradients over time.

\section{Summary}\label{sec:summary}

We characterized 19 star-forming regions throughout the local system NGC 1313 using spectroscopic data obtained with GMOS-S in multi-slit mode. We aimed to estimate and analyze the main physical properties of NGC 1313 and thus understand whether its perturbed morphology, associated with a possible interaction, has altered its evolution.

The regions observed in NGC 1313 are all H$\textsc{ii}$ regions, whose ionization is driven by massive O- and B-type stars through their UV radiation. Furthermore, these regions exhibit low electron densities ($n_e<$10 to 142 cm$^{-3}$),  consistent with typical H$\textsc{ii}$ regions in a barred spiral  galaxy such as NGC 1313, which is often considered transitional between the SMC and the LMC. The regions are also young and massive, with ages younger than 6 Myr and masses between $\sim$ 10$^{4.9}$ and 10$^{6.3}$ M$_\odot$. This is consistent with the results of \citet{finn+2024a}, who show that NGC 1313 is efficient at forming young and massive clusters.

The NGC 1313 system shows a low oxygen abundance, with metallicities ranging from 12+log(O/H) = 8.01 to 8.22, with a mean value of 12+log(O/H) = 8.13. These values place NGC 1313 in between the SMC and LMC in terms of metallicity. When estimating the abundance gradient, we find a nearly flat chemical gradient, regardless of the adjustment made to the regions. However, separating between the northeast and southwest regions shows that the southwestern region drives this slight overabundance in the outskirts of the galaxy, making the gradient slope slightly positive or at least flat when considering all the regions in a single fit. Based on this, gas flows become a likely scenario, where gas inflows or outflows may have diluted the central metallicity and spread enriched material, respectively.

Finally, we suggest an evolutionary scenario in which NGC 1313 interacted with a satellite galaxy in the southwestern region. This likely caused the flattening of the abundance gradient, the low chemical abundances and electron densities, and the presence of young, moderately massive regions. However, the gradient in the northeast arm is also interesting, as flattening is observed. This may indicate the presence of gas flows that mix chemical material throughout the galaxy.

NGC 1313 thus represents an ideal laboratory for studying how interactions affect particularly low-mass barred spiral galaxies. Its perturbed morphology suggests a past minor interaction with a satellite galaxy in the southwestern region of the system. This minor interaction left subtle traces, including a perturbed arm, a nearly flat abundance gradient, young massive H\textsc{ii} regions, and localized chemical enrichment in the southwestern region.

\begin{acknowledgements}

We thank the anonymous referee for reviewing this paper, providing comments that improved its scientific quality. BO-D and ST-F acknowledges the financial support of ULS/DIDULS through a regular project number PR2453858 and through an \textit{Apoyo de Tesis de Postgrado} project number PTE2314. CL-D acknowledges support from the ANID through Fondecyt project 3250511.

\end{acknowledgements}

\bibliographystyle{aa}
\bibliography{ref}

\begin{thebibliography}{80}
\expandafter\ifx\csname natexlab\endcsname\relax\def\natexlab#1{#1}\fi

\bibitem[{{Alfaro-Cuello} {et~al.}(2015){Alfaro-Cuello}, {Torres-Flores}, {Carrasco}, {Mendes de Oliveira}, {de Mello}, \& {Amram}}]{AlfaroCuello15}
{Alfaro-Cuello}, M., {Torres-Flores}, S., {Carrasco}, E.~R., {et~al.} 2015, \mnras, 453, 1355

\bibitem[{{Allende Prieto} {et~al.}(2001){Allende Prieto}, {Lambert}, \& {Asplund}}]{allende-prieto+2001}
{Allende Prieto}, C., {Lambert}, D.~L., \& {Asplund}, M. 2001, \apjl, 556, L63

\bibitem[{{Amram} {et~al.}(1996){Amram}, {Balkowski}, {Boulesteix}, {Cayatte}, {Marcelin}, \& {Sullivan}}]{amram+1996}
{Amram}, P., {Balkowski}, C., {Boulesteix}, J., {et~al.} 1996, \aap, 310, 737

\bibitem[{{Amram} {et~al.}(2007){Amram}, {Mendes de Oliveira}, {Plana}, {Balkowski}, \& {Hernandez}}]{Amram2007}
{Amram}, P., {Mendes de Oliveira}, C., {Plana}, H., {Balkowski}, C., \& {Hernandez}, O. 2007, \aap, 471, 753

\bibitem[{{Baldwin} {et~al.}(1981){Baldwin}, {Phillips}, \& {Terlevich}}]{bpt1981}
{Baldwin}, J.~A., {Phillips}, M.~M., \& {Terlevich}, R. 1981, \pasp, 93, 5

\bibitem[{{Barnes} \& {Hernquist}(1992)}]{barnes&hernquist1992}
{Barnes}, J.~E. \& {Hernquist}, L. 1992, \araa, 30, 705

\bibitem[{{Bresolin} {et~al.}(2012){Bresolin}, {Kennicutt}, \& {Ryan-Weber}}]{bresolin+2012}
{Bresolin}, F., {Kennicutt}, R.~C., \& {Ryan-Weber}, E. 2012, \apj, 750, 122

\bibitem[{{Bustamante} {et~al.}(2018){Bustamante}, {Sparre}, {Springel}, \& {Grand}}]{bustamante+2018}
{Bustamante}, S., {Sparre}, M., {Springel}, V., \& {Grand}, R. J.~J. 2018, \mnras, 479, 3381

\bibitem[{{Buzzo} {et~al.}(2021){Buzzo}, {Ziegler}, {Amram}, {Verdugo}, {Barbosa}, {Ciocan}, {Papaderos}, {Torres-Flores}, \& {Mendes de Oliveira}}]{buzzo+2021}
{Buzzo}, M.~L., {Ziegler}, B., {Amram}, P., {et~al.} 2021, \mnras, 503, 106

\bibitem[{{Calzetti} {et~al.}(2000){Calzetti}, {Armus}, {Bohlin}, {Kinney}, {Koornneef}, \& {Storchi-Bergmann}}]{calzetti+2000}
{Calzetti}, D., {Armus}, L., {Bohlin}, R.~C., {et~al.} 2000, \apj, 533, 682

\bibitem[{{Calzetti} {et~al.}(2015){Calzetti}, {Lee}, {Sabbi}, {Adamo}, {Smith}, {Andrews}, {Ubeda}, {Bright}, {Thilker}, {Aloisi}, {Brown}, {Chandar}, {Christian}, {Cignoni}, {Clayton}, {da Silva}, {de Mink}, {Dobbs}, {Elmegreen}, {Elmegreen}, {Evans}, {Fumagalli}, {Gallagher}, {Gouliermis}, {Grebel}, {Herrero}, {Hunter}, {Johnson}, {Kennicutt}, {Kim}, {Krumholz}, {Lennon}, {Levay}, {Martin}, {Nair}, {Nota}, {{\"O}stlin}, {Pellerin}, {Prieto}, {Regan}, {Ryon}, {Schaerer}, {Schiminovich}, {Tosi}, {Van Dyk}, {Walterbos}, {Whitmore}, \& {Wofford}}]{calzetti+2015}
{Calzetti}, D., {Lee}, J.~C., {Sabbi}, E., {et~al.} 2015, \aj, 149, 51

\bibitem[{{Chen} {et~al.}(2023){Chen}, {Grasha}, {Battisti}, {Kewley}, {Madore}, {Seibert}, {Rich}, \& {Beaton}}]{chen+2023}
{Chen}, Q.-H., {Grasha}, K., {Battisti}, A.~J., {et~al.} 2023, \mnras, 519, 4801

\bibitem[{{Cid Fernandes} {et~al.}(2011){Cid Fernandes}, {Stasi{\'n}ska}, {Mateus}, \& {Vale Asari}}]{cidfernandes+2011}
{Cid Fernandes}, R., {Stasi{\'n}ska}, G., {Mateus}, A., \& {Vale Asari}, N. 2011, \mnras, 413, 1687

\bibitem[{{Copetti} {et~al.}(2000){Copetti}, {Mallmann}, {Schmidt}, \& {Casta{\~n}eda}}]{copetti+2000}
{Copetti}, M.~V.~F., {Mallmann}, J.~A.~H., {Schmidt}, A.~A., \& {Casta{\~n}eda}, H.~O. 2000, \aap, 357, 621

\bibitem[{{De Robertis} {et~al.}(1987){De Robertis}, {Dufour}, \& {Hunt}}]{derobertis+1987}
{De Robertis}, M.~M., {Dufour}, R.~J., \& {Hunt}, R.~W. 1987, \jrasc, 81, 195

\bibitem[{{de Vaucouleurs}(1963)}]{deVaucouleurs1963}
{de Vaucouleurs}, G. 1963, \apj, 137, 720

\bibitem[{{Dey} {et~al.}(2019){Dey}, {Schlegel}, {Lang}, {Blum}, {Burleigh}, {Fan}, {Findlay}, {Finkbeiner}, {Herrera}, {Juneau}, {Landriau}, {Levi}, {McGreer}, {Meisner}, {Myers}, {Moustakas}, {Nugent}, {Patej}, {Schlafly}, {Walker}, {Valdes}, {Weaver}, {Y{\`e}che}, {Zou}, {Zhou}, {Abareshi}, {Abbott}, {Abolfathi}, {Aguilera}, {Alam}, {Allen}, {Alvarez}, {Annis}, {Ansarinejad}, {Aubert}, {Beechert}, {Bell}, {BenZvi}, {Beutler}, {Bielby}, {Bolton}, {Brice{\~n}o}, {Buckley-Geer}, {Butler}, {Calamida}, {Carlberg}, {Carter}, {Casas}, {Castander}, {Choi}, {Comparat}, {Cukanovaite}, {Delubac}, {DeVries}, {Dey}, {Dhungana}, {Dickinson}, {Ding}, {Donaldson}, {Duan}, {Duckworth}, {Eftekharzadeh}, {Eisenstein}, {Etourneau}, {Fagrelius}, {Farihi}, {Fitzpatrick}, {Font-Ribera}, {Fulmer}, {G{\"a}nsicke}, {Gaztanaga}, {George}, {Gerdes}, {Gontcho}, {Gorgoni}, {Green}, {Guy}, {Harmer}, {Hernandez}, {Honscheid}, {Huang}, {James}, {Jannuzi}, {Jiang}, {Joyce}, {Karcher}, {Karkar}, {Kehoe}, {Kneib}, {Kueter-Young}, {Lan},
  {Lauer}, {Le Guillou}, {Le Van Suu}, {Lee}, {Lesser}, {Perreault Levasseur}, {Li}, {Mann}, {Marshall}, {Mart{\'\i}nez-V{\'a}zquez}, {Martini}, {du Mas des Bourboux}, {McManus}, {Meier}, {M{\'e}nard}, {Metcalfe}, {Mu{\~n}oz-Guti{\'e}rrez}, {Najita}, {Napier}, {Narayan}, {Newman}, {Nie}, {Nord}, {Norman}, {Olsen}, {Paat}, {Palanque-Delabrouille}, {Peng}, {Poppett}, {Poremba}, {Prakash}, {Rabinowitz}, {Raichoor}, {Rezaie}, {Robertson}, {Roe}, {Ross}, {Ross}, {Rudnick}, {Safonova}, {Saha}, {S{\'a}nchez}, {Savary}, {Schweiker}, {Scott}, {Seo}, {Shan}, {Silva}, {Slepian}, {Soto}, {Sprayberry}, {Staten}, {Stillman}, {Stupak}, {Summers}, {Sien Tie}, {Tirado}, {Vargas-Maga{\~n}a}, {Vivas}, {Wechsler}, {Williams}, {Yang}, {Yang}, {Yapici}, {Zaritsky}, {Zenteno}, {Zhang}, {Zhang}, {Zhou}, \& {Zhou}}]{dey+2019}
{Dey}, A., {Schlegel}, D.~J., {Lang}, D., {et~al.} 2019, \aj, 157, 168

\bibitem[{{Dom{\'\i}nguez} {et~al.}(2013){Dom{\'\i}nguez}, {Siana}, {Henry}, {Scarlata}, {Bedregal}, {Malkan}, {Atek}, {Ross}, {Colbert}, {Teplitz}, {Rafelski}, {McCarthy}, {Bunker}, {Hathi}, {Dressler}, {Martin}, \& {Masters}}]{dominguez+2013}
{Dom{\'\i}nguez}, A., {Siana}, B., {Henry}, A.~L., {et~al.} 2013, \apj, 763, 145

\bibitem[{{Doran} {et~al.}(2013){Doran}, {Crowther}, {de Koter}, {Evans}, {McEvoy}, {Walborn}, {Bastian}, {Bestenlehner}, {Gr{\"a}fener}, {Herrero}, {K{\"o}hler}, {Ma{\'\i}z Apell{\'a}niz}, {Najarro}, {Puls}, {Sana}, {Schneider}, {Taylor}, {van Loon}, \& {Vink}}]{doran+2013}
{Doran}, E.~I., {Crowther}, P.~A., {de Koter}, A., {et~al.} 2013, \aap, 558, A134

\bibitem[{{Draine}(2011)}]{draine2011}
{Draine}, B.~T. 2011, {Physics of the Interstellar and Intergalactic Medium}

\bibitem[{{Ellison} {et~al.}(2008){Ellison}, {Patton}, {Simard}, \& {McConnachie}}]{ellison+2008}
{Ellison}, S.~L., {Patton}, D.~R., {Simard}, L., \& {McConnachie}, A.~W. 2008, \aj, 135, 1877

\bibitem[{{Fern{\'a}ndez} {et~al.}(2024){Fern{\'a}ndez}, {Amor{\'\i}n}, {Firpo}, \& {Morisset}}]{fernandez+2024}
{Fern{\'a}ndez}, V., {Amor{\'\i}n}, R., {Firpo}, V., \& {Morisset}, C. 2024, \aap, 688, A69

\bibitem[{{Finn} {et~al.}(2024){Finn}, {Johnson}, {Indebetouw}, {Costa}, {Adamo}, {Aloisi}, {Bittle}, {Calzetti}, {Dale}, {Dobbs}, {Donovan Meyer}, {Elmegreen}, {Elmegreen}, {Fumagalli}, {Gallagher}, {Grasha}, {Grebel}, {Kennicutt}, {Krumholz}, {Lee}, {Messa}, {Nair}, {Sabbi}, {Smith}, {Thilker}, {Whitmore}, \& {Wofford}}]{finn+2024a}
{Finn}, M.~K., {Johnson}, K.~E., {Indebetouw}, R., {et~al.} 2024, \apj, 964, 12

\bibitem[{{Flaugher} {et~al.}(2015){Flaugher}, {Diehl}, {Honscheid}, {Abbott}, {Alvarez}, {Angstadt}, {Annis}, {Antonik}, {Ballester}, {Beaufore}, {Bernstein}, {Bernstein}, {Bigelow}, {Bonati}, {Boprie}, {Brooks}, {Buckley-Geer}, {Campa}, {Cardiel-Sas}, {Castander}, {Castilla}, {Cease}, {Cela-Ruiz}, {Chappa}, {Chi}, {Cooper}, {da Costa}, {Dede}, {Derylo}, {DePoy}, {de Vicente}, {Doel}, {Drlica-Wagner}, {Eiting}, {Elliott}, {Emes}, {Estrada}, {Fausti Neto}, {Finley}, {Flores}, {Frieman}, {Gerdes}, {Gladders}, {Gregory}, {Gutierrez}, {Hao}, {Holland}, {Holm}, {Huffman}, {Jackson}, {James}, {Jonas}, {Karcher}, {Karliner}, {Kent}, {Kessler}, {Kozlovsky}, {Kron}, {Kubik}, {Kuehn}, {Kuhlmann}, {Kuk}, {Lahav}, {Lathrop}, {Lee}, {Levi}, {Lewis}, {Li}, {Mandrichenko}, {Marshall}, {Martinez}, {Merritt}, {Miquel}, {Mu{\~n}oz}, {Neilsen}, {Nichol}, {Nord}, {Ogando}, {Olsen}, {Palaio}, {Patton}, {Peoples}, {Plazas}, {Rauch}, {Reil}, {Rheault}, {Roe}, {Rogers}, {Roodman}, {Sanchez}, {Scarpine}, {Schindler}, {Schmidt},
  {Schmitt}, {Schubnell}, {Schultz}, {Schurter}, {Scott}, {Serrano}, {Shaw}, {Smith}, {Soares-Santos}, {Stefanik}, {Stuermer}, {Suchyta}, {Sypniewski}, {Tarle}, {Thaler}, {Tighe}, {Tran}, {Tucker}, {Walker}, {Wang}, {Watson}, {Weaverdyck}, {Wester}, {Woods}, {Yanny}, \& {DES Collaboration}}]{flaugher+2015}
{Flaugher}, B., {Diehl}, H.~T., {Honscheid}, K., {et~al.} 2015, \aj, 150, 150

\bibitem[{{Friedli} {et~al.}(1994){Friedli}, {Benz}, \& {Kennicutt}}]{friedli+1994}
{Friedli}, D., {Benz}, W., \& {Kennicutt}, R. 1994, \apjl, 430, L105

\bibitem[{{G{\'o}mez-Espinoza} {et~al.}(2023){G{\'o}mez-Espinoza}, {Torres-Flores}, {Firpo}, {Amram}, {Epinat}, {Contini}, \& {Mendes de Oliveira}}]{GomezEspinoza2023}
{G{\'o}mez-Espinoza}, D.~A., {Torres-Flores}, S., {Firpo}, V., {et~al.} 2023, \mnras, 522, 2655

\bibitem[{{Grand} {et~al.}(2012){Grand}, {Kawata}, \& {Cropper}}]{grand+2012}
{Grand}, R. J.~J., {Kawata}, D., \& {Cropper}, M. 2012, \mnras, 426, 167

\bibitem[{{Grand} {et~al.}(2016){Grand}, {Springel}, {Kawata}, {Minchev}, {S{\'a}nchez-Bl{\'a}zquez}, {G{\'o}mez}, {Marinacci}, {Pakmor}, \& {Campbell}}]{grand+2016}
{Grand}, R. J.~J., {Springel}, V., {Kawata}, D., {et~al.} 2016, \mnras, 460, L94

\bibitem[{{Gray} \& {Corbally}(2009)}]{gray+2009}
{Gray}, R.~O. \& {Corbally}, J., C. 2009, {Stellar Spectral Classification}

\bibitem[{{Hadfield} \& {Crowther}(2007)}]{hadfield+2007}
{Hadfield}, L.~J. \& {Crowther}, P.~A. 2007, \mnras, 381, 418

\bibitem[{{Hannon} {et~al.}(2019){Hannon}, {Lee}, {Whitmore}, {Chandar}, {Adamo}, {Mobasher}, {Aloisi}, {Calzetti}, {Cignoni}, {Cook}, {Dale}, {Deger}, {Della Bruna}, {Elmegreen}, {Gouliermis}, {Grasha}, {Grebel}, {Herrero}, {Hunter}, {Johnson}, {Kennicutt}, {Kim}, {Sacchi}, {Smith}, {Thilker}, {Turner}, {Walterbos}, \& {Wofford}}]{hannon+2019}
{Hannon}, S., {Lee}, J.~C., {Whitmore}, B.~C., {et~al.} 2019, \mnras, 490, 4648

\bibitem[{{Hernandez} {et~al.}(2022){Hernandez}, {Winch}, {Larsen}, {James}, \& {Jones}}]{Hernandez2022}
{Hernandez}, S., {Winch}, A., {Larsen}, S., {James}, B.~L., \& {Jones}, L. 2022, \aj, 164, 89

\bibitem[{{Iglesias-P{\'a}ramo} {et~al.}(2006){Iglesias-P{\'a}ramo}, {Buat}, {Takeuchi}, {Xu}, {Boissier}, {Boselli}, {Burgarella}, {Madore}, {Gil de Paz}, {Bianchi}, {Barlow}, {Byun}, {Donas}, {Forster}, {Friedman}, {Heckman}, {Jelinski}, {Lee}, {Malina}, {Martin}, {Milliard}, {Morrissey}, {Neff}, {Rich}, {Schiminovich}, {Seibert}, {Siegmund}, {Small}, {Szalay}, {Welsh}, \& {Wyder}}]{iglesias-paramo2006}
{Iglesias-P{\'a}ramo}, J., {Buat}, V., {Takeuchi}, T.~T., {et~al.} 2006, \apjs, 164, 38

\bibitem[{{Jin} {et~al.}(2023){Jin}, {Sutherland}, {Kewley}, \& {Nicholls}}]{jin+2023}
{Jin}, Y., {Sutherland}, R., {Kewley}, L.~J., \& {Nicholls}, D.~C. 2023, \apj, 958, 179

\bibitem[{{Kauffmann} {et~al.}(2003){Kauffmann}, {Heckman}, {Tremonti}, {Brinchmann}, {Charlot}, {White}, {Ridgway}, {Brinkmann}, {Fukugita}, {Hall}, {Ivezi{\'c}}, {Richards}, \& {Schneider}}]{kauffmann+2003}
{Kauffmann}, G., {Heckman}, T.~M., {Tremonti}, C., {et~al.} 2003, \mnras, 346, 1055

\bibitem[{{Kaviraj}(2014)}]{kaviraj+2014}
{Kaviraj}, S. 2014, \mnras, 440, 2944

\bibitem[{{Kennicutt}(1998)}]{kennicutt1998}
{Kennicutt}, Jr., R.~C. 1998, \araa, 36, 189

\bibitem[{{Kewley} {et~al.}(2006){Kewley}, {Groves}, {Kauffmann}, \& {Heckman}}]{kewley+2006}
{Kewley}, L.~J., {Groves}, B., {Kauffmann}, G., \& {Heckman}, T. 2006, \mnras, 372, 961

\bibitem[{{Kewley} {et~al.}(2001){Kewley}, {Heisler}, {Dopita}, \& {Lumsden}}]{kewley+2001}
{Kewley}, L.~J., {Heisler}, C.~A., {Dopita}, M.~A., \& {Lumsden}, S. 2001, \apjs, 132, 37

\bibitem[{{Kewley} {et~al.}(2010){Kewley}, {Rupke}, {Zahid}, {Geller}, \& {Barton}}]{kewley+2010}
{Kewley}, L.~J., {Rupke}, D., {Zahid}, H.~J., {Geller}, M.~J., \& {Barton}, E.~J. 2010, \apjl, 721, L48

\bibitem[{{Krabbe} {et~al.}(2014){Krabbe}, {Rosa}, {Dors}, {Pastoriza}, {Winge}, {H{\"a}gele}, {Cardaci}, \& {Rodrigues}}]{krabbe+2014}
{Krabbe}, A.~C., {Rosa}, D.~A., {Dors}, O.~L., {et~al.} 2014, \mnras, 437, 1155

\bibitem[{{Kroupa} {et~al.}(2024){Kroupa}, {Gjergo}, {Jerabkova}, \& {Yan}}]{kroupa+2024}
{Kroupa}, P., {Gjergo}, E., {Jerabkova}, T., \& {Yan}, Z. 2024, arXiv e-prints, arXiv:2410.07311

\bibitem[{{Leitherer} {et~al.}(1999){Leitherer}, {Schaerer}, {Goldader}, {Delgado}, {Robert}, {Kune}, {de Mello}, {Devost}, \& {Heckman}}]{leitherer+1999}
{Leitherer}, C., {Schaerer}, D., {Goldader}, J.~D., {et~al.} 1999, \apjs, 123, 3

\bibitem[{{L{\'o}pez-S{\'a}nchez} {et~al.}(2015){L{\'o}pez-S{\'a}nchez}, {Westmeier}, {Esteban}, \& {Koribalski}}]{lopez-sanchez+2015}
{L{\'o}pez-S{\'a}nchez}, {\'A}.~R., {Westmeier}, T., {Esteban}, C., \& {Koribalski}, B.~S. 2015, \mnras, 450, 3381

\bibitem[{{Maeder}(1992)}]{maeder1992}
{Maeder}, A. 1992, \aap, 264, 105

\bibitem[{{Magrini} {et~al.}(2017){Magrini}, {Gon{\c{c}}alves}, \& {Vajgel}}]{magrini+2017}
{Magrini}, L., {Gon{\c{c}}alves}, D.~R., \& {Vajgel}, B. 2017, \mnras, 464, 739

\bibitem[{{Marino} {et~al.}(2013){Marino}, {Rosales-Ortega}, {S{\'a}nchez}, {Gil de Paz}, {V{\'\i}lchez}, {Miralles-Caballero}, {Kehrig}, {P{\'e}rez-Montero}, {Stanishev}, {Iglesias-P{\'a}ramo}, {D{\'\i}az}, {Castillo-Morales}, {Kennicutt}, {L{\'o}pez-S{\'a}nchez}, {Galbany}, {Garc{\'\i}a-Benito}, {Mast}, {Mendez-Abreu}, {Monreal-Ibero}, {Husemann}, {Walcher}, {Garc{\'\i}a-Lorenzo}, {Masegosa}, {Del Olmo Orozco}, {Mour{\~a}o}, {Ziegler}, {Moll{\'a}}, {Papaderos}, {S{\'a}nchez-Bl{\'a}zquez}, {Gonz{\'a}lez Delgado}, {Falc{\'o}n-Barroso}, {Roth}, {van de Ven}, \& {CALIFA Team}}]{marino+2013}
{Marino}, R.~A., {Rosales-Ortega}, F.~F., {S{\'a}nchez}, S.~F., {et~al.} 2013, \aap, 559, A114

\bibitem[{{McGaugh} {et~al.}(2000){McGaugh}, {Schombert}, {Bothun}, \& {de Blok}}]{2000McGaugh}
{McGaugh}, S.~S., {Schombert}, J.~M., {Bothun}, G.~D., \& {de Blok}, W.~J.~G. 2000, \apjl, 533, L99

\bibitem[{{Medoff} {et~al.}(2025){Medoff}, {Mutlu-Pakdil}, {Carlin}, {Drlica-Wagner}, {Tollerud}, {Doliva-Dolinsky}, {Sand}, {Mart{\'\i}nez-V{\'a}zquez}, {Stringfellow}, {Cerny}, {Crnojevi{\'c}}, {Ferguson}, {Fielder}, {Chaturvedi}, {Kallivayalil}, {No{\"e}l}, {Vivas}, {Walker}, {Adam{\'o}w}, {Bom}, {Carballo-Bello}, {Choi}, {Medina}, {Navabi}, {Pace}, {Riley}, \& {Sakowska}}]{medoff+2025}
{Medoff}, J., {Mutlu-Pakdil}, B., {Carlin}, J.~L., {et~al.} 2025, \apj, 990, 108

\bibitem[{{Mendes de Oliveira} {et~al.}(2017){Mendes de Oliveira}, {Amram}, {Quint}, {Torres-Flores}, {Barb{\'a}}, \& {Andrade}}]{2017MendesdeOliveira}
{Mendes de Oliveira}, C., {Amram}, P., {Quint}, B.~C., {et~al.} 2017, \mnras, 469, 3424

\bibitem[{{Mu{\~n}oz-Elgueta} {et~al.}(2018){Mu{\~n}oz-Elgueta}, {Torres-Flores}, {Amram}, {Hernandez-Jimenez}, {Urrutia-Viscarra}, {Mendes de Oliveira}, \& {G{\'o}mez-L{\'o}pez}}]{munoz-elgueta+2018}
{Mu{\~n}oz-Elgueta}, N., {Torres-Flores}, S., {Amram}, P., {et~al.} 2018, \mnras, 480, 3257

\bibitem[{{Ohlson} {et~al.}(2024){Ohlson}, {Seth}, {Gallo}, {Baldassare}, \& {Greene}}]{ohlson+2024}
{Ohlson}, D., {Seth}, A.~C., {Gallo}, E., {Baldassare}, V.~F., \& {Greene}, J.~E. 2024, \aj, 167, 31

\bibitem[{{Olave-Rojas} {et~al.}(2015){Olave-Rojas}, {Torres-Flores}, {Carrasco}, {Mendes de Oliveira}, {de Mello}, \& {Scarano}}]{olave-rojas+2015}
{Olave-Rojas}, D., {Torres-Flores}, S., {Carrasco}, E.~R., {et~al.} 2015, \mnras, 453, 2808

\bibitem[{{Osterbrock} \& {Ferland}(2006)}]{OF2006}
{Osterbrock}, D.~E. \& {Ferland}, G.~J. 2006, {Astrophysics of gaseous nebulae and active galactic nuclei}

\bibitem[{{Pagel} {et~al.}(1980){Pagel}, {Edmunds}, \& {Smith}}]{pagel+1980}
{Pagel}, B.~E.~J., {Edmunds}, M.~G., \& {Smith}, G. 1980, \mnras, 193, 219

\bibitem[{{Pan} {et~al.}(2025){Pan}, {Lin}, {S{\'a}nchez}, {Barrera-Ballesteros}, \& {Hsieh}}]{pan+2025}
{Pan}, H.-A., {Lin}, L., {S{\'a}nchez}, S.~F., {Barrera-Ballesteros}, J.~K., \& {Hsieh}, B.-C. 2025, \apj, 982, 130

\bibitem[{{Papovich} {et~al.}(2015){Papovich}, {Labb{\'e}}, {Quadri}, {Tilvi}, {Behroozi}, {Bell}, {Glazebrook}, {Spitler}, {Straatman}, {Tran}, {Cowley}, {Dav{\'e}}, {Dekel}, {Dickinson}, {Ferguson}, {Finkelstein}, {Gawiser}, {Inami}, {Faber}, {Kacprzak}, {Kawinwanichakij}, {Kocevski}, {Koekemoer}, {Koo}, {Kurczynski}, {Lotz}, {Lu}, {Lucas}, {McIntosh}, {Mehrtens}, {Mobasher}, {Monson}, {Morrison}, {Nanayakkara}, {Persson}, {Salmon}, {Simons}, {Tomczak}, {van Dokkum}, {Weiner}, \& {Willner}}]{papovich+2015}
{Papovich}, C., {Labb{\'e}}, I., {Quadri}, R., {et~al.} 2015, \apj, 803, 26

\bibitem[{{Patton} {et~al.}(2013){Patton}, {Torrey}, {Ellison}, {Mendel}, \& {Scudder}}]{patton+2013}
{Patton}, D.~R., {Torrey}, P., {Ellison}, S.~L., {Mendel}, J.~T., \& {Scudder}, J.~M. 2013, \mnras, 433, L59

\bibitem[{{Patton} {et~al.}(2020){Patton}, {Wilson}, {Metrow}, {Ellison}, {Torrey}, {Brown}, {Hani}, {McAlpine}, {Moreno}, \& {Woo}}]{patton+2020}
{Patton}, D.~R., {Wilson}, K.~D., {Metrow}, C.~J., {et~al.} 2020, \mnras, 494, 4969

\bibitem[{{Perez} {et~al.}(2011){Perez}, {Michel-Dansac}, \& {Tissera}}]{perez+2011}
{Perez}, J., {Michel-Dansac}, L., \& {Tissera}, P.~B. 2011, \mnras, 417, 580

\bibitem[{{Peters} {et~al.}(1994){Peters}, {Freeman}, {Forster}, {Manchester}, \& {Ables}}]{peters+1994}
{Peters}, W.~L., {Freeman}, K.~C., {Forster}, J.~R., {Manchester}, R.~N., \& {Ables}, J.~G. 1994, \mnras, 269, 1025

\bibitem[{{Portinari} {et~al.}(2004){Portinari}, {Sommer-Larsen}, \& {Tantalo}}]{2004Portinari}
{Portinari}, L., {Sommer-Larsen}, J., \& {Tantalo}, R. 2004, \mnras, 347, 691

\bibitem[{{Rosa} {et~al.}(2014){Rosa}, {Dors}, {Krabbe}, {H{\"a}gele}, {Cardaci}, {Pastoriza}, {Rodrigues}, \& {Winge}}]{rosa+2014}
{Rosa}, D.~A., {Dors}, O.~L., {Krabbe}, A.~C., {et~al.} 2014, \mnras, 444, 2005

\bibitem[{{Rupke} {et~al.}(2010){Rupke}, {Kewley}, \& {Barnes}}]{rupke+2010a}
{Rupke}, D. S.~N., {Kewley}, L.~J., \& {Barnes}, J.~E. 2010, \apjl, 710, L156

\bibitem[{{Russell} \& {Dopita}(1990)}]{russell+1990}
{Russell}, S.~C. \& {Dopita}, M.~A. 1990, \apjs, 74, 93

\bibitem[{{Ryder} {et~al.}(1993){Ryder}, {Staveley-Smith}, {Dopita}, {Petre}, {Colbert}, {Malin}, \& {Schlegel}}]{ryder+1993}
{Ryder}, S., {Staveley-Smith}, L., {Dopita}, M., {et~al.} 1993, \apj, 416, 167

\bibitem[{{Ryder} {et~al.}(1995){Ryder}, {Staveley-Smith}, {Malin}, \& {Walsh}}]{ryder+1995}
{Ryder}, S.~D., {Staveley-Smith}, L., {Malin}, D., \& {Walsh}, W. 1995, \aj, 109, 1592

\bibitem[{{Sabbi} {et~al.}(2018){Sabbi}, {Calzetti}, {Ubeda}, {Adamo}, {Cignoni}, {Thilker}, {Aloisi}, {Elmegreen}, {Elmegreen}, {Gouliermis}, {Grebel}, {Messa}, {Smith}, {Tosi}, {Dolphin}, {Andrews}, {Ashworth}, {Bright}, {Brown}, {Chandar}, {Christian}, {Clayton}, {Cook}, {Dale}, {de Mink}, {Dobbs}, {Evans}, {Fumagalli}, {Gallagher}, {Grasha}, {Herrero}, {Hunter}, {Johnson}, {Kahre}, {Kennicutt}, {Kim}, {Krumholz}, {Lee}, {Lennon}, {Martin}, {Nair}, {Nota}, {{\"O}stlin}, {Pellerin}, {Prieto}, {Regan}, {Ryon}, {Sacchi}, {Schaerer}, {Schiminovich}, {Shabani}, {Van Dyk}, {Walterbos}, {Whitmore}, \& {Wofford}}]{sabbi+2018}
{Sabbi}, E., {Calzetti}, D., {Ubeda}, L., {et~al.} 2018, \apjs, 235, 23

\bibitem[{{Scarano} {et~al.}(2008){Scarano}, {Madsen}, {Roy}, \& {L{\'e}pine}}]{scarano+2008}
{Scarano}, S., {Madsen}, F. R.~H., {Roy}, N., \& {L{\'e}pine}, J.~R.~D. 2008, \mnras, 386, 963

\bibitem[{{Silva-Villa} \& {Larsen}(2012)}]{silva-villa+2012}
{Silva-Villa}, E. \& {Larsen}, S.~S. 2012, \mnras, 423, 213

\bibitem[{{Suzuki} {et~al.}(2013){Suzuki}, {Kaneda}, \& {Onaka}}]{suzuki+2013}
{Suzuki}, T., {Kaneda}, H., \& {Onaka}, T. 2013, \aap, 554, A8

\bibitem[{{Tapia-Contreras} {et~al.}(2025){Tapia-Contreras}, {Tissera}, {Sillero}, {Gonzalez-Jara}, {Casanueva-Villarreal}, {Pedrosa}, {Bignone}, {Padilla}, \& {Dom{\'\i}nguez-Tenreiro}}]{tapia-contreras+2025}
{Tapia-Contreras}, B., {Tissera}, P.~B., {Sillero}, E., {et~al.} 2025, \aap, 700, A69

\bibitem[{{Taylor} {et~al.}(2011){Taylor}, {Hopkins}, {Baldry}, {Brown}, {Driver}, {Kelvin}, {Hill}, {Robotham}, {Bland-Hawthorn}, {Jones}, {Sharp}, {Thomas}, {Liske}, {Loveday}, {Norberg}, {Peacock}, {Bamford}, {Brough}, {Colless}, {Cameron}, {Conselice}, {Croom}, {Frenk}, {Gunawardhana}, {Kuijken}, {Nichol}, {Parkinson}, {Phillipps}, {Pimbblet}, {Popescu}, {Prescott}, {Sutherland}, {Tuffs}, {van Kampen}, \& {Wijesinghe}}]{taylor+2011}
{Taylor}, E.~N., {Hopkins}, A.~M., {Baldry}, I.~K., {et~al.} 2011, \mnras, 418, 1587

\bibitem[{{Tokovinin} {et~al.}(2008){Tokovinin}, {Tighe}, {Schurter}, {Cantarutti}, {van der Bliek}, {Martinez}, {Mondaca}, \& {Montane}}]{2008Tokovinin}
{Tokovinin}, A., {Tighe}, R., {Schurter}, P., {et~al.} 2008, in Society of Photo-Optical Instrumentation Engineers (SPIE) Conference Series, Vol. 7015, Adaptive Optics Systems, ed. N.~{Hubin}, C.~E. {Max}, \& P.~L. {Wizinowich}, 70154C

\bibitem[{{Toomre} \& {Toomre}(1972)}]{toomre&toomre1972}
{Toomre}, A. \& {Toomre}, J. 1972, \apj, 178, 623

\bibitem[{{Toribio San Cipriano} {et~al.}(2017){Toribio San Cipriano}, {Dom{\'\i}nguez-Guzm{\'a}n}, {Esteban}, {Garc{\'\i}a-Rojas}, {Mesa-Delgado}, {Bresolin}, {Rodr{\'\i}guez}, \& {Sim{\'o}n-D{\'\i}az}}]{toribio_san_cipriano+2017}
{Toribio San Cipriano}, L., {Dom{\'\i}nguez-Guzm{\'a}n}, G., {Esteban}, C., {et~al.} 2017, \mnras, 467, 3759

\bibitem[{{Torres-Flores} {et~al.}(2014){Torres-Flores}, {Scarano}, {Mendes de Oliveira}, {de Mello}, {Amram}, \& {Plana}}]{torres-flores+2014}
{Torres-Flores}, S., {Scarano}, S., {Mendes de Oliveira}, C., {et~al.} 2014, \mnras, 438, 1894

\bibitem[{{van Dokkum}(2001)}]{vanDokkum2001}
{van Dokkum}, P.~G. 2001, \pasp, 113, 1420

\bibitem[{{Walsh} \& {Roy}(1997)}]{walsh+1997}
{Walsh}, J.~R. \& {Roy}, J.~R. 1997, \mnras, 288, 726

\bibitem[{{Whitmore} {et~al.}(2023){Whitmore}, {Chandar}, {Rodr{\'\i}guez}, {Lee}, {Emsellem}, {Floyd}, {Kim}, {Kruijssen}, {Mok}, {Sormani}, {Boquien}, {Dale}, {Faesi}, {Henny}, {Hannon}, {Thilker}, {White}, {Barnes}, {Bigiel}, {Chevance}, {Henshaw}, {Klessen}, {Leroy}, {Liu}, {Maschmann}, {Meidt}, {Rosolowsky}, {Schinnerer}, {Sun}, {Watkins}, \& {Williams}}]{whitmore+2023}
{Whitmore}, B.~C., {Chandar}, R., {Rodr{\'\i}guez}, M.~J., {et~al.} 2023, \apjl, 944, L14

\end{thebibliography}

\end{document}